\documentclass[]{aa}

\usepackage{graphicx}
\usepackage{txfonts}
\usepackage{indentfirst}
\usepackage{natbib}
\usepackage{textcomp}
\usepackage{ulem}
\bibpunct{(}{)}{;}{a}{}{,} 
\usepackage[scriptsize]{subfigure}
\usepackage{amsmath}

\usepackage[colorlinks=true]{hyperref}
\usepackage{color}

\newcommand{\msun}{$\mathrm{M}_\odot$}

\begin{document}
  
  \title{Discs and outflows in the early phases of massive star formation: influence of magnetic fields and ambipolar diffusion}
   \author{B. Commerçon
           \inst{1}
           ,
	 M. González\inst{2},
	 R. Mignon-Risse\inst{3},
	 P. Hennebelle\inst{2}
 	  \and
            N. Vaytet\inst{4}
          }
   \offprints{B. Commerçon}

   \institute{Univ Lyon, Ens de Lyon, Univ Lyon 1, CNRS, Centre de Recherche Astrophysique de Lyon UMR5574, 69007, Lyon, France\\
             \email{benoit.commercon@ens-lyon.fr}
              \and
              AIM, CEA, CNRS, Universit\'e Paris-Saclay, Universit\'e de Paris, 91191 Gif-sur-Yvette, France
	\and
	Universit\'e de Paris, CNRS, AstroParticule et Cosmologie, F-75013 Paris, France
	\and
Data Management and Software Centre, European Spallation Source ERIC, Ole Maaløes Vej 3, 2200, Copenhagen, Denmark           }

   \date {...; ...}

  \abstract  {Massive star formation remains one of the most challenging problems in astrophysics, as illustrated by the fundamental issues of the radiative pressure barrier and the initial fragmentation. The wide variety of physical processes involved, in particular the protostellar radiative feedback, increases the complexity of massive star formation in comparison with its low-mass counterpart.}   
  {We aim to study the details of mass accretion and ejection in the vicinity of massive star forming cores using high-resolution (5~au) 3-dimensional numerical simulations. We investigate the mechanisms at the origin of outflows (radiative force versus magnetic acceleration). We characterise the properties of the disc forming around massive protostars depending on the physics included: hydrodynamics, magnetic fields, and ambipolar diffusion.}
  {We use state-of-the-art 3-dimensional adaptive-mesh-refinement models of massive dense cores collapse which integrate the equations of (resistive) grey radiation-magnetohydrodynamics, and include sink particles evolution. For the first time, we include both protostellar radiative feedback via pre-main sequence evolutionary tracks and magnetic ambipolar diffusion. To determine the role of magnetic fields and AD play in the formation of outflows and discs, we studied 3 different cases: a purely hydrodynamical run, a magnetised simulation under the ideal approximation (perfect coupling), and a calculation with ambipolar diffusion (resistive case). In the most micro-physically complex model (resistive MHD), we also investigated the effect the initial amplitude of both magnetic field and solid body rotation have on the final properties of the massive protostellar system. We use simple criteria to identify the outflow and disc material and follow their evolution as the central star accretes mass up to $~20$~\msun~ in most of our models. The radiative, magnetic, and hydrodynamical properties of the outflows and discs are quantitatively measured and cross compared-between models.}
  {Massive stars form in all our models, together with outflows and discs. The outflow is completely different when magnetic fields are introduced, so that magneto-centrifugal processes are the main driver of  the outflow up to stellar masses of $~20$~\msun. Then, the disc properties heavily depend on the physics included. In particular, the disc formed in the ideal and resistive runs show opposite properties in terms of plasma beta, i.e. the ratio of thermal to magnetic pressures, and of magnetic fields topology. While the disc in the ideal case is dominated by the magnetic pressure and the toroidal magnetic fields, the one formed in the resistive runs is dominated by the thermal pressure and has essentially a vertical magnetic fields in the inner regions ($R<100-200$~au). }
  {We find that magnetic processes dominate the early evolution of massive protostellar systems ($M_\star<20$~\msun)  and shapes the accretion/ejection as well as the disc formation. Ambipolar diffusion is mainly at work at disc scales and regulates its properties. We  predict magnetic fields topology within the disc and outflows, as well as disc masses and radii to be compared with observations. Last, our finding for the outflow and disc properties are reminiscent of the low-mass star formation framework, suggesting that accretion and ejection in young massive and low-mass protostars are regulated by the same physical processes at the early stages.}


\titlerunning{Discs and outflows in the early phases of massive star formation}
\authorrunning{B. Commer\c con et al.}
   \maketitle


\section{Introduction}

Massive stars dominate from birth to death the energy input in the interstellar medium. As such, their formation mechanism needs to be better understood in order to put constraints on their impact on star formation and galaxy evolution as a whole. 
It is currently admitted that massive stars form in giant molecular clouds, which exhibit turbulent motions and magnetic fields \citep[e.g.][]{tan:14,motte:17}. The mechanisms by which massive stars get their mass in this giant, turbulent, and magnetised complex remain poorly constrained though. In particular, two major issues make it difficult to assemble a large reservoir of mass within a single massive star: the radiative pressure barrier and the initial fragmentation. 
Thanks to the formidable increase of observational capabilities, it is now possible to probe massive star forming regions with an unprecedented angular resolution corresponding to sub-50~au scales \citep[e.g.][]{beuther:19,maud:19}. In parallel, heavy numerical developments are undertaken in order to describe accurately the physics of star formation in numerical models \citep[see for a recent review][]{teyssier:19}. 

A number of observational  studies focus on the fragmentation of massive cores which are thought to be the progenitors of massive stars \citep[e.g.][]{bontemps:10,palau:13,zhang:15,fontani:16,csengeri:17,figueira18,sanhueza:19,sanna:19}. There is no clear consensus on which physical processes is regulating the fragmentation level. There is however a trend in which massive dense cores are not as fragmented as expected \citep{wang:14,csengeri:17,nony:18}, and that magnetic fields has an important role in the fragmentation \citep{li:15,dallolio:19}. Using state-of-the-art multi-dimensional and multi-physics numerical models, it has also been demonstrated that magnetic fields, radiative transfer and turbulence all play role in setting the fragmentation level of massive star forming regions \citep{hennebelle:11,commercon:11b,myers:13,fontani:18}. It is worth to note that both, radiative feedback and magnetic fields can be very efficient at preventing massive dense cores to fragment heavily \citep{commercon:11b,myers:13}.

At core scales, important efforts are done to investigate the mechanisms which regulate mass accretion and ejection in the vicinity of massive star forming regions. There are growing evidences that massive stars form in a similar way to low-mass stars \cite[e.g.][]{zhang:19,beltran:14,tan:16} and that discs and outflows are formed in the early stages \citep{duarte:13,tan:16,yang:18,zhang:19}. In addition, an increasing number of studies reports the detection of an hourglass magnetic field at massive core scales  \citep{girart:09,qiu:14,ching:16,beltran:19}. In particular, \cite{qiu:14}  estimate a magnetic field strength of $\sim1.1$~mG, resulting in a mass-to-magnetic flux ratio of 1.4 times the critical value. A mG magnetic field amplitude is consistent with estimates in other massive star forming region \citep[e.g.][]{girart:13,dallolio:19,beltran:19}. Last but not least, \cite{girart:09} and \cite{qiu:14} conclude that their observations are consistent with magnetic braking, i.e. a magnetically regulated collapse. 

A first direct observation of a disc around a young massive protostar was reported in \cite{kraus:10}, where they find evidences in IRAS 13481-6124 of a very compact disc ($< 20$~au) at the late stages, i.e. after the main accretion phase. However, direct evidences for discs in young, deeply embedded sources is expected to be weak  because of confusion with the surrounding envelope \citep{cesaroni:17}. \cite{girart:17,girart:18} present subarcsecond angular resolution observations  with the Submillimeter Array (SMA) and the Atacama Large Millimeter Array (ALMA) interferometers of the B0-type protostar GGD27 MM1, which is powering the HH80-81 radio jet. Their observations clearly resolve a disc oriented perpendicularly to the radio jet, with a radius of about 300~au. \cite{beuther:19} presented sub-50~au scales observation of the hot core region G351.77-0.54 with ALMA. They do not report the presence of a large ($R>1000$ au) disc. Additional spatially resolved compact discs are also reported in the recent literature \citep{fernandez:16,motogi:19}. Multidimensional numerical simulations have also investigated disc formation and evolution in massive dense core collapse, either in the hydrodynamical case \citep[e.g.][]{yorke:02,klassen:16,kuiper:18} or in the ideal magneto-hydrodynamics (MHD) case \citep[e.g.][]{banerjee:06,seifried:13,myers:13,gray:18}. Pure hydrodynamical and ideal MHD models, represent the two extreme limits of the coupling between the gas and magnetic fields. Indeed, non-ideal coupling effects such as the Ohmic diffusion, the Hall effect and the ambipolar diffusion are at play in the physical conditions typical of protostellar collapse \citep[e.g.][]{marchand:16}. Contrary to low-mass star formation \citep[e.g][]{masson:16,hennebelle:20}, the effect of non-ideal MHD has not been investigated extensively in the massive star formation framework \citep[but see][]{matsushita:17,kolligan:18}.

In low-mass protostars, it is now admitted that the observed bipolar outflows and jets are launched via magneto-centrifugal processes \citep{franck:14,li:14b}. Recently, \cite{ching:16} reported SMA observations toward NGC 1333 IRAS 4A of CO polarization and find evidence of helical magnetic fields in the outflow.   Outflows and jets are also observed  in high-mass star forming regions \citep{shepherd:96, zhang:01,beuther:02,duarte:13,maud:15,mcleod:18}. There are  observational evidences that these outflows and jets are linked with magnetic fields.  First, synchrotron emission has been associated with jets in observations of massive protostellar sources \citep{carasco:10,beltran:16b,sanna:19b,zhang:19b}. Second,  observations of an alignment between outflows and magnetic fields direction inferred from spectral-line linear polarization have also been reported in massive star forming regions \citep{cortes:06,beuther:10,girart:13}. 
Importantly, in the high-mass regime, the protostellar luminosity becomes high enough to balance the gravity and prevent further accretion: the radiative pressure barrier. Multidimensional numerical models have shown that indeed the radiative force may accelerate the gas and generate also bipolar outflows that channel radiation to escape to allow accretion through the disc \citep{yorke:02,kuiper:12,klassen:16} or via radiative Rayleigh-Taylor instabilities  \citep{krumholz:07a,rosen:16}. Using sub-grid models, \cite{kuiper:15} include both the radiative feedback and protostellar outflow from massive stars and show that protostellar outflows help to form more massive stars. \cite{seifried:12} using ideal MHD and \cite{matsushita:17} using resistive MHD performed 3D collapse simulations of magnetised massive dense cores and reported outflow formation self-consistently driven by magneto centrifugal processes, but they neglected protostellar radiative feedback.   
There is to date no study that explores the combined effect of magnetic fields and protostellar radiative feedback on the launching of outflows in massive protostellar systems.

The aim of this paper is to lay the foundations of  the combined effects of protostellar radiative feedback and of magnetic fields in the formation of the star-disc-outflow system in massive collapsing dense cores. This work is part of a more general framework of numerical experiments, in which we increase step-by-step the complexity of the physics included. As it is now widely done in the low-mass regime, we introduce resistive effects through ambipolar diffusion in order to take into account the non-perfect coupling of gas with magnetic fields. We also integrate the effect of protostellar radiative feedback using sub-grid pre-main sequence evolutionary tracks to compute the protostellar luminosity. We use 3D dimensional dynamical models of massive dense core protostellar collapse,  introducing gradually more realistic  physical processes (magnetic fields, ambipolar diffusion).   In this study, we use a grey approximation for the protostellar irradiation, which is known to underestimate the radiative acceleration  \citep[e.g.][]{kuiper:10,mignon:20}. In \cite{mignon:20}, we present an hybrid method which is built over the grey irradiation scheme and shows good performances in estimating the radiative acceleration. First results, focusing on the interplay between turbulence, ambipolar diffusion and accurate stellar irradiation are presented in a related study \citep{mignon:21a,mignon:21b}.

The paper is organised as follows. In section \ref{Sec:methods} we detail our numerical method, setup, and initial conditions. We present our results in Sect.~\ref{Sec:results}, with a particular focus on the outflows and discs properties. Section \ref{Sec:discussion} is devoted to highlighting the role of ambipolar diffusion in the disc formation process, and to the comparison with observational and other numerical works.  We also discuss in Sect.~\ref{Sec:discussion} the limits of our work and propose future direction of studies. Section \ref{Sec:conclusion} concludes our work.

\section{Methods and initial conditions \label{Sec:methods}}
\subsection{Radiation magneto-hydrodynamics model}

Our numerical model integrates the equation of radiation-magneto-hydrodynamics (RMHD) and includes radiative protostellar feedback. We detail in the following our numerical tool. 

We use the adaptive-mesh-refinement (AMR) code \ttfamily{RAMSES}\rm~\citep{teyssier:02} which is based on a finite-volume second-order Godunov scheme and a constrained transport algorithm for ideal magnetohydrodynamics \citep{fromang:06,teyssier:06}. The non-ideal MHD solver, including the effect of ambipolar diffusion and Ohmic diffusion as corrections of the electromagnetic force (EMF) at the cell edges, is presented in \cite{masson:12}. In this study, we account only for the ambipolar diffusion in the non-ideal MHD runs. In addition, we use the radiation-hydrodynamics (RHD) solver  presented in a series of papers \citep{commercon:11a,commercon:14,gonzalez:15}, which integrates the equation of RHD in the flux-limited diffusion approximation \citep[FLD, e.g.][]{minerbo:78,levermore:81}. The full RMHD equations with all the radiative quantities estimated in the co-moving frame and under the grey approximation (radiative quantities are integrated over the entire frequency spectrum)  read
 \begin{equation}
\begin{array}{llll}
\partial_t \rho + \nabla\cdot \left[\rho\textbf{u} \right] & =  0, \\
\partial_t \rho \textbf{u} + \nabla \cdot\left[\rho \textbf{u}\otimes \textbf{u} + P \mathbb{I} \right]& = - \lambda\nabla E_\mathrm{r} +\textbf{F}_\mathrm{L} -\rho\nabla\phi,\\
\partial_t E_\mathrm{T} + \nabla\cdot \left[\textbf{u}\left( E_\mathrm{T} + P_\mathrm{} + \frac{B^2}{2} \right)  \right.&  \\
\hspace{30pt}\left.-\left(\textbf{u}\cdot\textbf{B}\right)\textbf{B}-\textbf{E}_\mathrm{AD}\times\textbf{B}\right] &=  - \mathbb{P}_\mathrm{r}\nabla:\textbf{u}  - \lambda \textbf{u} \nabla E_\mathrm{r} \\
 &   \hspace{9pt} +  \nabla \cdot\left(\frac{c\lambda}{\rho \kappa_\mathrm{R}} \nabla E_\mathrm{r}\right)+L_\star\\&\hspace{9pt} -\rho\textbf{u}\cdot\nabla\phi, \\
\partial_t E_\mathrm{r} + \nabla\cdot \left[\textbf{u}E_\mathrm{r}\right]
&= 
- \mathbb{P}_\mathrm{r}\nabla:\textbf{u}  +  \nabla \cdot\left(\frac{c\lambda}{\rho \kappa_\mathrm{R}} \nabla E_\mathrm{r}\right) \\
 &  \hspace{9pt} + \kappa_\mathrm{P}\rho c(a_\mathrm{R}T^4 - E_\mathrm{r})\\
 &\hspace{9pt} +L_\star,\\ 
\partial_t \textbf{B} - \nabla\times\left[\textbf{u}\times\textbf{B} +\textbf{E}_\mathrm{AD}\right]&=0,\\
\nabla\cdot\textbf{B}&=0,\\
\Delta\phi & = 4\pi G \rho,
\end{array}
\end{equation}
\noindent where $\rho$ is the material density, $\textbf{u}$ is the velocity, $P$ the thermal pressure,  $\lambda$ is the radiative flux limiter, $E_\mathrm{r}$ is the radiative
 energy, $\textbf{F}_\mathrm{L}=(\nabla\times\textbf{B})\times\textbf{B}$ is the Lorentz force, $\phi$ is the gravitational potential, $E_\mathrm{T}$ the total energy $E_\mathrm{T}=\rho\epsilon +1/2\rho u^2 + 1/2B^2+E_\mathrm{r}$ ($\epsilon$ is the gas internal specific energy), $\textbf{B}$ is the magnetic field, $\textbf{E}_\mathrm{AD}$ is the ambipolar EMF, $\kappa_\mathrm{P}$ is the Planck mean opacity,  $\kappa_
\mathrm{R}$ is the Rosseland mean opacity, $\mathbb{P}_\mathrm{r}$ is the radiation pressure, $L_\star$ the luminosity source (i.e., protostar luminosity)
, and $T$ is the gas temperature. 
The system is closed using the perfect gas relation, $P=\rho k_\mathrm{B}T/(\mu_\mathrm{gas} m_\mathrm{H})$, with  $\mu_\mathrm{gas}$ the gas mean molecular weight.

The ambipolar EMF is given by 
\begin{equation}
\textbf{E}_\mathrm{AD} = \frac{\eta_\mathrm{AD}}{B^2}\left[\left(\nabla\times\textbf{B}\right) \times \textbf{B} \right] \times \textbf{B},
\end{equation}
where $\eta_\mathrm{AD}$ is the ambipolar diffusion resistivity, calculated as a function of the density, temperature, and magnetic field amplitude. We use the abundances from the equilibrium chemistry code described in \cite{marchand:16}, which depends on the density and temperature.  As noted by \cite{shu:92} and \cite{masson:12}, the ambipolar diffusion, i.e. the neutral-ion friction, results in a  heat source term in the gas internal energy equation
\begin{equation}
\partial_t{\rho\epsilon}=\Gamma_\mathrm{AD} = -\textbf{E}_\mathrm{AD} \cdot \left(\nabla \times \textbf{B} \right) =\frac{\eta_\mathrm{AD}}{B^2}\left\Vert\left(\nabla \times \textbf{B} \right) \times\textbf{B}\right\Vert^2.
\label{eq:heating}
\end{equation}

For the radiative transfer, we use the grey Rosseland and Planck opacities tabulated in \cite{vaytet:13}, who compiled the dust opacity from \cite{semenov:03}, the molecular gas opacity form \cite{ferguson:05}, and the atomic gas opacity from \cite{badnell:05}.

\subsection{Initial conditions}

Our initial conditions are similar to the ones found in previous works focusing on isolated massive star formation \citep[e.g.][]{krumholz:09,kuiper:10,klassen:16}. 
We consider 100 \msun~spherical dense cores with an  uniform temperature of 20~K. The initial density profile is centrally condensed and follows $\rho(r)=\rho_\mathrm{c}/(1+(r/r_\mathrm{c})^{-2})$, where the central density is $\rho_\mathrm{c}\sim 7.7\times10^{-18}$~g~cm$^{-3}$ (corresponding to a freefall time of~24~kyr) and the density contrast between the centre and the border of the core equals 100. The extent of the central plateau is $r_\mathrm{c}=0.02$~pc and the initial core radius is $r_0=0.2$~pc. We use an ideal gas equation of state, with the mean molecular weight $\mu_\mathrm{gas}=2.31$ and the specific heat ratio $\gamma=5/3$. This corresponds to a ratio of thermal to gravitational energies of 6\%.
We impose an initial solid-body rotation along the $x$-axis with angular frequency (resp. ratio of rotational to gravitational energies $\beta$) $\Omega_\mathrm{s}\approx 9.5\times 10^{-15}$~Hz ($\beta \approx0.01$, slow rotation), and  $\Omega_\mathrm{f}\approx  2.1\times10^{-14}$~Hz ($\beta \approx0.05$, fast rotation).
We deliberately choose a low angular velocity to foster on the star-disc-outflow system formation without undergoing heavy fragmentation in the process. We do not apply any initial perturbation in the different RMHD variables. 

The magnetic field is initially aligned with the rotation axis and is uniform through the cloud. Its strength is given by the mass-to-flux parameter $\mu$, which represents the ratio between the mass-to-flux calculated at the border of the core $(M/\phi)_0=M_0/(\pi B_0r_0^2)$ and the critical mass-to-flux ratio $(M/\phi)_\mathrm{crit}=0.53/(3\pi)(5/G)^{1/2}$ \citep{mouschovias:76}. This choice of initial magnetic configuration makes the mass-to-flux ratio $\mu$ not uniform with radius, scaling as $\mu \sim 1/R$, up to a value $\mu_\mathrm{c}$ of 10 times the initial one in the central plateau region, which represents a relatively weak magnetization (e.g. $\mu_\mathrm{c}=50$ for   $\mu=5$).

\subsection{Numerical resolution and sink particles}
The simulation box is four times larger than the initial core radius ($\sim 0.8$ pc). The coarser grid resolution is $64^3$ and we allow for 9 additional levels of refinement, which gives a minimum resolution of $\Delta x_\mathrm{min}=5$~au. The grid is refined using a Jeans length criterion with at least 10 points per Jeans length. We use periodic boundary conditions\footnote{We run a comparison simulations using isolated boundary conditions for the HYDRO case and our qualitative results remain unaffected. The accretion rate yet  increases, but only by less than maximum 15\% for the sink mass and it decreases with time (2.5\% at the time when the sink mass is 10 \msun.}.

We use sink particles  to describe the dynamics of the collapse below the maximum resolution. We use the publicly available implementation of sink particles in the \ttfamily{RAMSES}\rm~code \citep{bleuler:14}, with some modifications in the checks performed for their creation. We use the clump finder algorithm of \cite{bleuler:14} for the sink creation sites identification, but we do not apply their virial check. Instead, we perform  two alternative checks which are used in other popular implementations of sink particles \citep[e.g.][]{bate:95,federrath:10}: the bound state check, i.e. the total energy in the clump is negative ($E_\mathrm{pot} + E_\mathrm{kin}+E_\mathrm{therm}+ E_\mathrm{mag}+E_\mathrm{rad}<0$), and the Jeans instability check, i.e. the mass inside the clump should exceed the local Jeans mass ($E_\mathrm{pot}+2E_\mathrm{therm}<0$). We use a threshold density of $10^{10}$ cm$^{-3}$ for the clump identification. 

The accretion radius $r_\mathrm{acc}$ is set to $4\Delta x_\mathrm{min}\sim 20$~au, and we impose that the sink accretion volume sits at the maximum level of refinement. We do not allow sink particles creation if the density peak of the clump finder algorithm falls within the sink accretion radius of another sink.  The sink particles with mass $<1$ \msun~are evolved using the particle-mesh Cloud-in-Cell method of  \ttfamily{RAMSES}\rm~\citep{teyssier:02}, while the trajectory of sink particles more massive than 1 \msun~are computed using direct force summation \citep[both for the gas-sink and the sink-sink interactions][]{bleuler:14}. 

For the sink accretion, we use the threshold accretion scheme \citep{federrath:10,bleuler:14}. We scan the cells which centre lies within $r_\mathrm{acc}$ and check for Jeans unstable gas by computing the Jeans density in each cell $\rho_\mathrm{sink}$. The mass accreted by the sink particles from the cell is set to
\begin{equation}
\Delta m = \mathrm{max}(0.25(\rho-\rho_\mathrm{sink})(\Delta x_\mathrm{min})^3,0).
\end{equation}
It is not clear at this stage whether sink particles should accrete angular momentum or not. The subsequent evolution of the angular momentum transport inside the sink accretion radius cannot be followed in our models. We thus assume another sub-grid prescription for angular momentum. We define $f_\mathrm{acc,mom}$ as the fraction of angular momentum accreted by the sink. If $f_\mathrm{acc,mom}=1$, all the angular momentum is accreted by the sink particular. On the opposite, all the angular momentum stays onto the grid cell for $f_\mathrm{acc,mom}=0$. We choose $f_\mathrm{acc,mom}=1$ as the fiducial value in this study, and we present a comparison between these two extreme cases in Appendix \ref{Sec:appendix_faccmom}. Last but not least, magnetic fields are not accreted with the gas by the sink particle. 

Last, we force sink merging when two sinks sit in a common accretion volume, without any additional check other than the proximity criterion. This aggressive merging procedure favours the formation of more massive stars. 

\subsection{Sub-grid model for radiative feedback}
\label{sec:model}

The protostellar luminosity sources $L_\star$ are associated to the evolution of the sink particles. We use the pre-main sequence (PMS) evolution models of \cite{kuiper:13} to compute the protostellar properties (luminosity, radius). The PMS tracks have been obtained with the STELLAR evolution code \citep{Bodenheimer:07,yorke:08} and tabulated as a function of the stellar mass and the mass accretion rate. 

We assume that each sink particle represents a single protostar, and that {\it all} the mass accreted by the sink particle goes into the stellar content (most favourable case for the radiative feedback), i.e. $M_\mathrm{sink}=M_\star$.  In this work, we only include the effect of the protostellar internal luminosity and we neglect the accretion luminosity.  This choice is matter of debate of course, but since we are interested in the regime where the stellar mass gets larger than 8~\msun, it is fair to include only the internal luminosity which dominates in this mass range \citep[e.g.][]{hosokawa:09}. We discuss this limitation in Sect.~\ref{Sec:discussion_limits} and postpone the exploration of the accretion luminosity influence in future studies.
	  In order to get the internal luminosity from the PMS track tables,  we do not account for the instantaneous accretion rate, but we use the mean accretion rate defined as $\dot{M}_\mathrm{mean} = M_\star/t_\star$, where $M_\star$ and $t_\star$ are the mass and the age of the protostar.  \cite{kuiper:13} have shown that using the  mean accretion rate instead of the instantaneous one provides a reasonable	estimate of the protostar’s influence on their environment. The energy input $L_\star\Delta t$ is then spread uniformly over the sink accretion volume at each finer level timestep. We have tested the use of weighting functions for the energy input as in \cite{krumholz:07a}, and we found no significant differences in the results compared with the uniform input.

\subsection{Disc and outflow identification \label{disk_criteria}}

In this paper, we focus on the results after the first sink particle creation. We identify the disc and outflow regions given the global flow properties and the mass and position of the sink particles.  We observe in the simulations that sink particles can move back and forth around the centre of mass of the star-disc-outflow system ($\sim10$ au). These little oscillations cause troubles when estimating the velocity in the frame of the sink particles. We therefore neglect the sink velocity in our analysis.  
For the disc identification, we compute the angular momentum vector direction within a sphere of 100~au around the sink particle position. Then we estimate the velocity of the gas in the cylindrical coordinate system where the $z$-axis is aligned along the angular momentum vector. We apply the criteria derived in \cite{joos:12} to select the cells which constitute the rotationally supported disc. We briefly recall the criteria
\begin{itemize}
\item the density verifies $n>10^{9}$~cm$^{-3}$;
\item the gas is rotationally supported, $v_\phi>f_\mathrm{thres} v_r$;
\item the gas is close to hydrostatic equilibrium, $v_\phi>f_\mathrm{thres} v_z$;
\item the rotation support is larger than the thermal support,  \hspace{10pt}$\frac12 \rho v_\phi^2>f_\mathrm{thres} P$.
\end{itemize}
We use $f_\mathrm{thres}=2$. We note that we do not use the connectivity criterion of \cite{joos:12}, since we find by experience that it does not affect the mass of disc (and thus the radius in which most of the mass is contained). The disc radius is computed as the radius at which 90\% of the disc mass is contained.

 We then identify the outflow region as follows: we compute the radial velocity relative to the sink particle position and account for the cells which have positive radial velocity $v_r$ larger than the escape velocity $v_\mathrm{esc}=(2GM_\mathrm{sink}/r)^{1/2}$, where $r$ is the distance between the cell and the sink particle position. 

\begin{table}
	\caption{Summary of the simulations parameters (the star labels our fiducial model). }
	\centering
	\begin{tabular}{c c c c c}
		\hline \hline
		Model 						& $\mu$   & AD 		 & $\Omega_0$ (Hz) & $t_0$ (kyr)\\
		\hline
		HYDRO  			  & $\infty$ &     no    & $9.5\times 10^{-15}$ & 	28.0	  \\
		MU5I       			&      5       &     no    & $9.5\times 10^{-15}$ &	27.9		 \\
		MU2AD 			 &      2       &    yes    & $9.5\times 10^{-15}$ &	28.2		 \\
		$^*$MU5AD  	  &      5       &    yes    & $9.5\times 10^{-15}$ & 	27.9		\\
		MU5ADf 			&      5       &    yes    & $2.1\times 10^{-14}$ &		28.4	 \\
		\hline
	\end{tabular}
	\label{tab:models}
	\tablefoot{$\mu$ is the magnetisation parameter, AD means ambipolar diffusion, $\Omega_0$ is the initial rotation frequency, and $t_0$ is the time of the formation of the first sink particle.}
\end{table}

\begin{table*}[t]
	\caption{Summary of the different component mass. }
	\centering
	\begin{tabular}{c c c c c c c c c c c c}
		\hline \hline
		
		Component & \multicolumn{2}{c}{}&\multicolumn{3}{c}{Sink} &  \multicolumn{3}{c}{Disc} &  \multicolumn{3}{c}{Outflow} \\
		& $t_1$&$t_2$&$M(t_1$) & $M(t_2)$ & $\dot{M}$ & $M(t_1$) & $M(t_2)$ & $\dot{M}$& $M(t_1$) & $M(t_2)$ & $\dot{M}$\\
		\hline
		HYDRO  & 43.4 & 53.8 & 5.5 & 10.6 & $4.9\times10^{-4}$& 4.6 & 7.2 & $2.5\times10^{-4}$ & - & - & - \\
		MU5I   & 44.1& 61.1 & 5.0 & 10.3 & $3.1\times10^{-4}$ & 1.9 & 3.5 & $9.4\times10^{-5}$ & 1.2 & 1.1 & $-6.0\times10^{-6}$    \\
		MU2AD  & 41.9 & 61.1 & 5.0 & 10.0 & $2.6\times10^{-4}$& 1.8& 2.4 & $7.3\times10^{-5}$ & $4.6\times10^{-1}$ & 2.3 & $9.6\times10^{-5}$\\
		MU5AD  & 42.1 & 56.6 & 5.0 & 10.0 & $3.4\times10^{-4}$& 1.9   & 3.3 & $9.7\times10^{-5}$&1& 3.1 &$1.4\times10^{-4}$\\
		MU5ADf & 48.9 & 87 & 5.0 &  10.0 & $1.3\times10^{-4}$ &  3.1 & 4.4 & $3.4\times10^{-5}$&2.1& 9.2 &$1.9\times10^{-4}$\\
		\hline
	\end{tabular}
	\label{tab:mass}
		\tablefoot{Mass are given in \msun. mass accretion rates $\dot{M}$ in \msun/yr, and times in kyr. Time $t_1$ corresponds to the output time closest to $M_*=5$~\msun~and $t_2$ to $M_*=10$~\msun.  }
\end{table*}

\subsection{Simulations parameters}
The five models we present in the main part of this paper are summarised in Table \ref{tab:models}.   Our fiducial run (MU5AD) includes magnetic fields with a moderate intensity ($\mu=5$) and ambipolar diffusion, as well as a slow initial rotation. We then compare this fiducial case with the results obtained  with a hydrodynamical model (HYDRO), a magnetised model with ideal MHD (MU5I), a strongly magnetised ($\mu=2$) model with ambipolar diffusion (MU2AD), and a magnetised model  with ambipolar diffusion and a fast initial rotation (MU5ADf).  The HYDRO model (without magnetic fields) is very similar to what has been done in previous studies \citep[e.g.][]{krumholz:09,klassen:16} and will be a reference model to which we will compare with the literature. 

We also present in appendix \ref{Sec:appendix_resolution} a numerical convergence study based on our fiducial model and in appendix \ref{Sec:appendix_faccmom} a comparison of our fiducial model with a sub-grid accretion scheme in which all the angular momentum of the gas accreted is transferred to the sink particle $f_\mathrm{acc,mom}=1$.

\section{Results\label{Sec:results}}

\subsection{Overview}

\begin{figure*}[t]
	\includegraphics[width=0.5\textwidth]{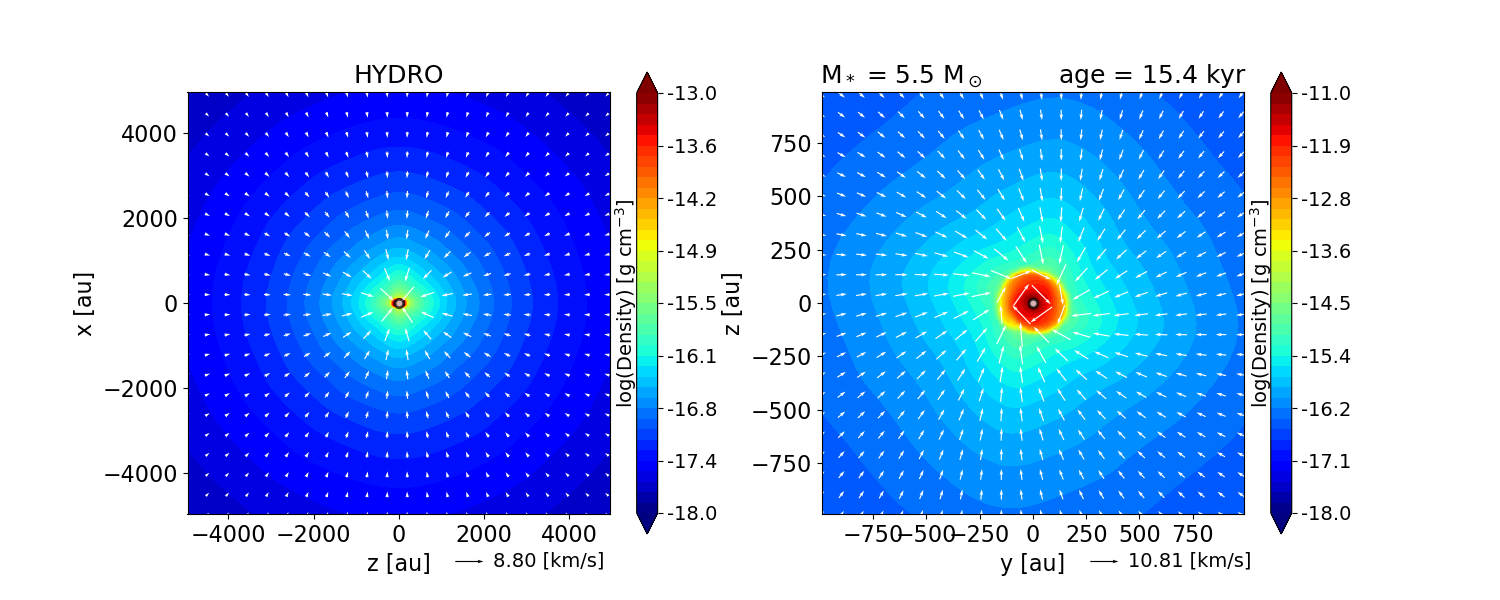}
	\includegraphics[width=0.5\textwidth]{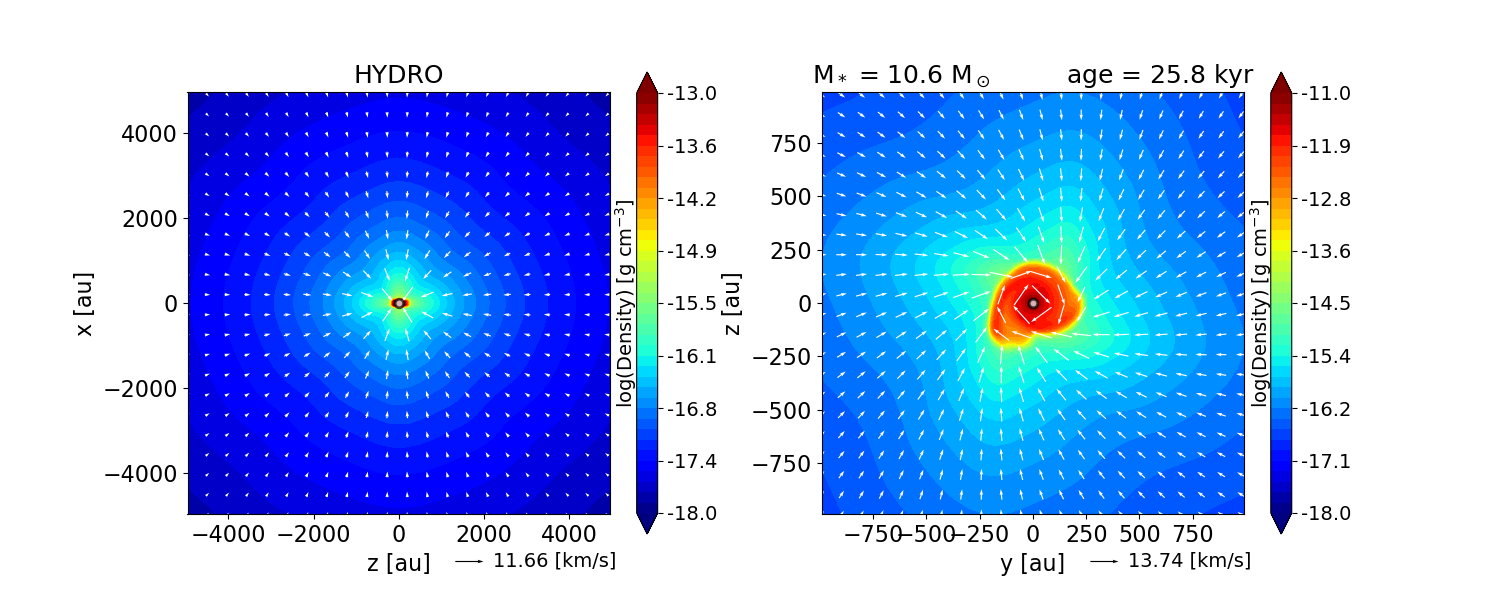}
	\includegraphics[width=0.5\textwidth]{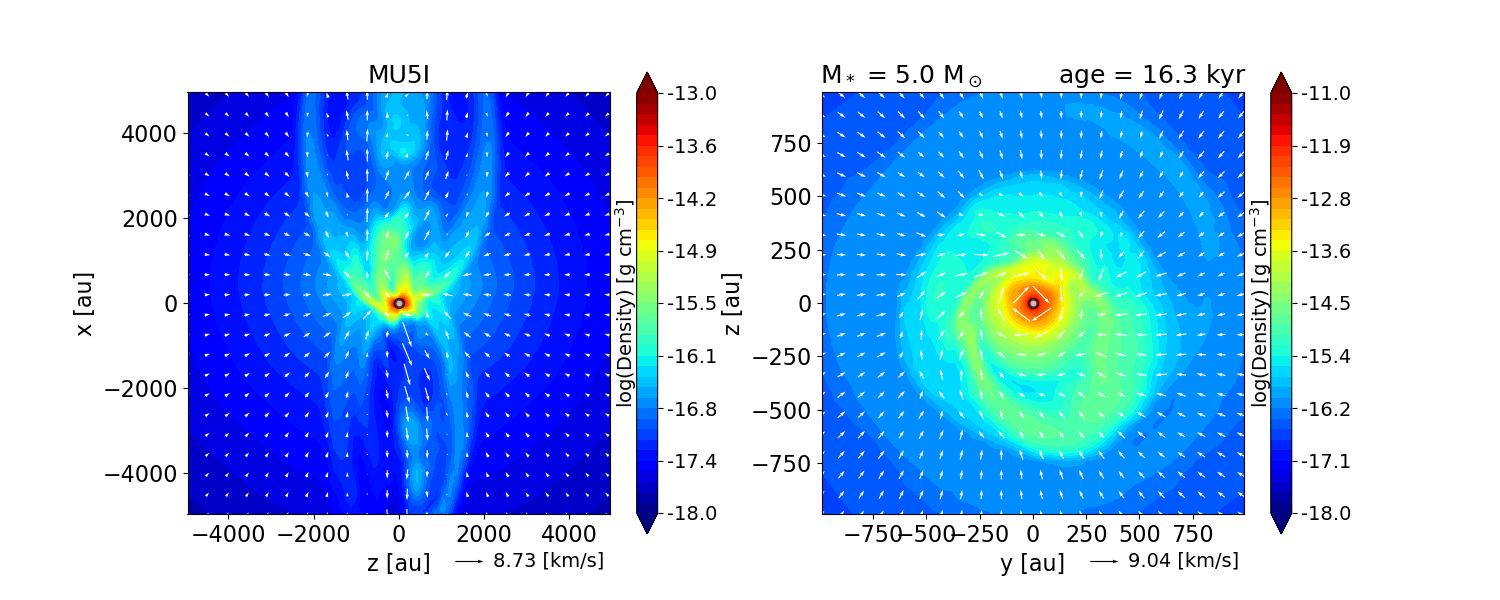}
	\includegraphics[width=0.5\textwidth]{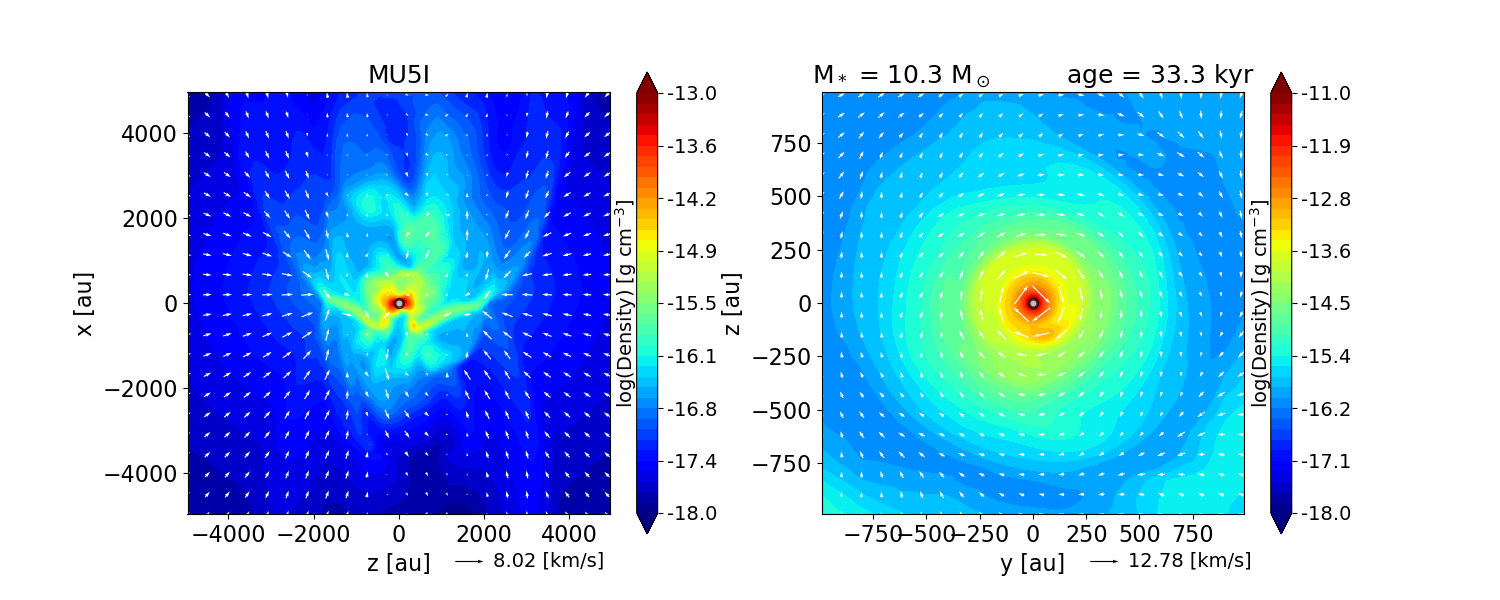}
	\includegraphics[width=0.5\textwidth]{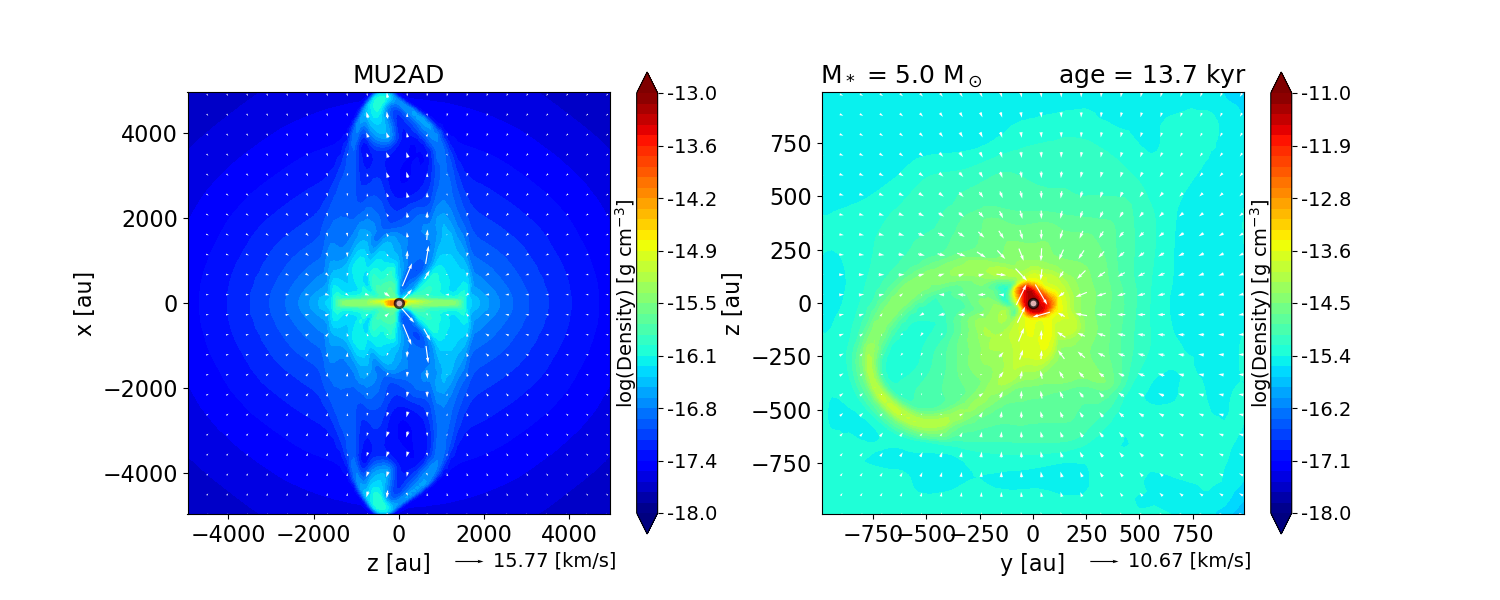}
	\includegraphics[width=0.5\textwidth]{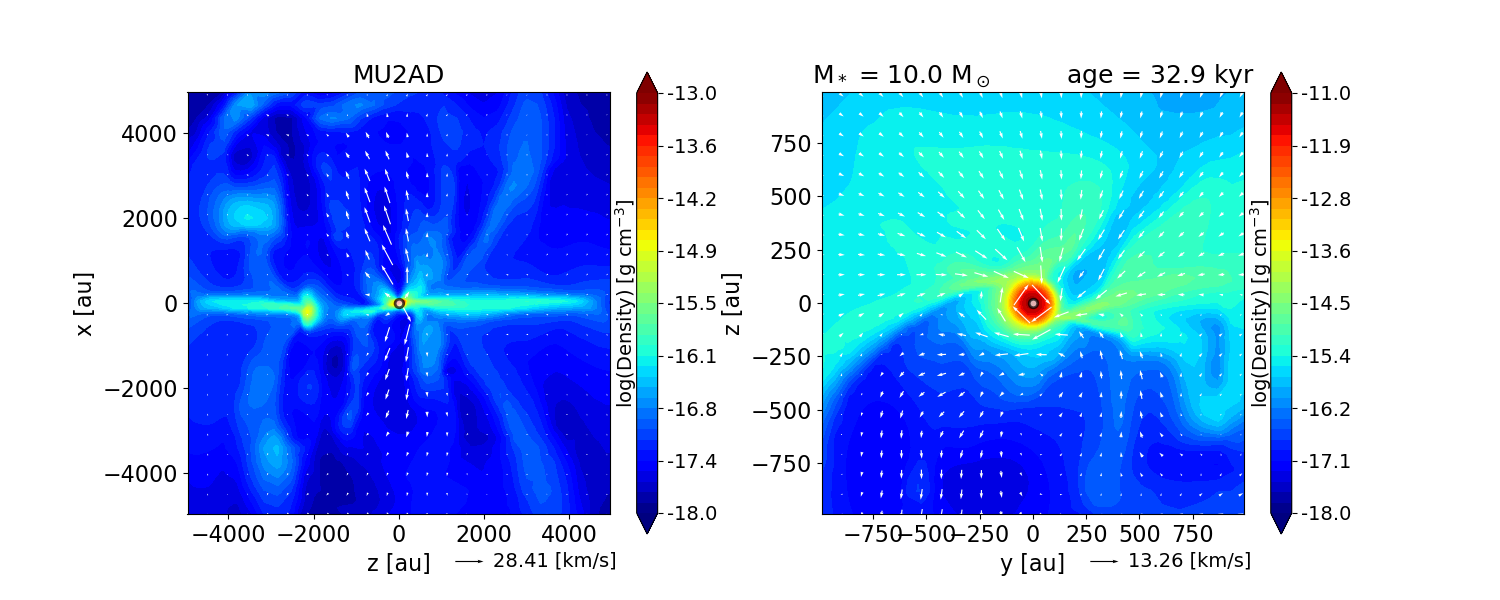}
	\includegraphics[width=0.5\textwidth]{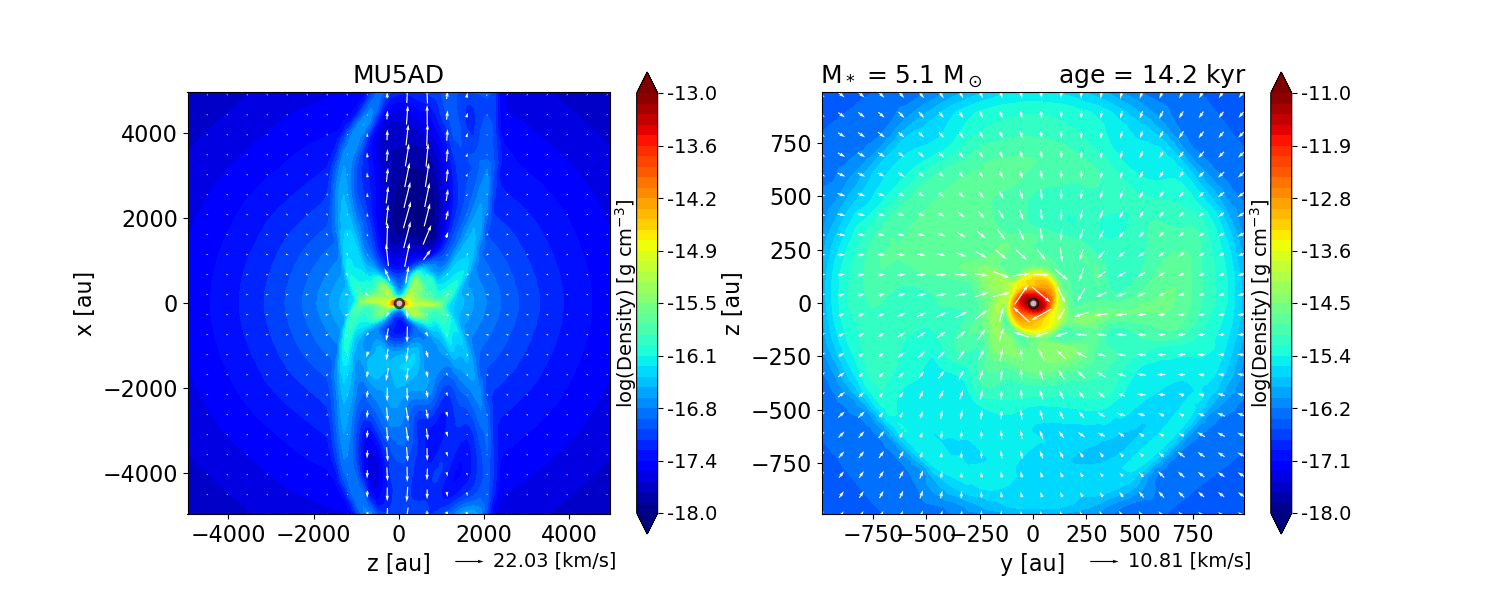}
	\includegraphics[width=0.5\textwidth]{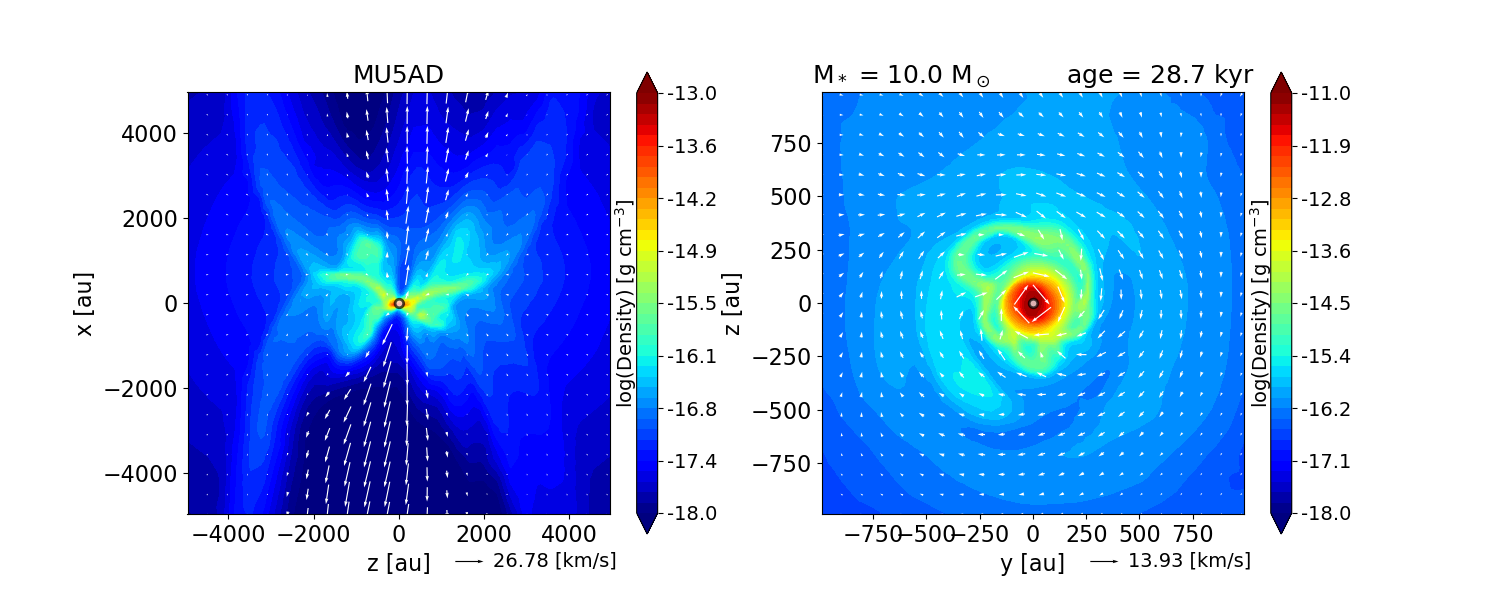}
	\includegraphics[width=0.5\textwidth]{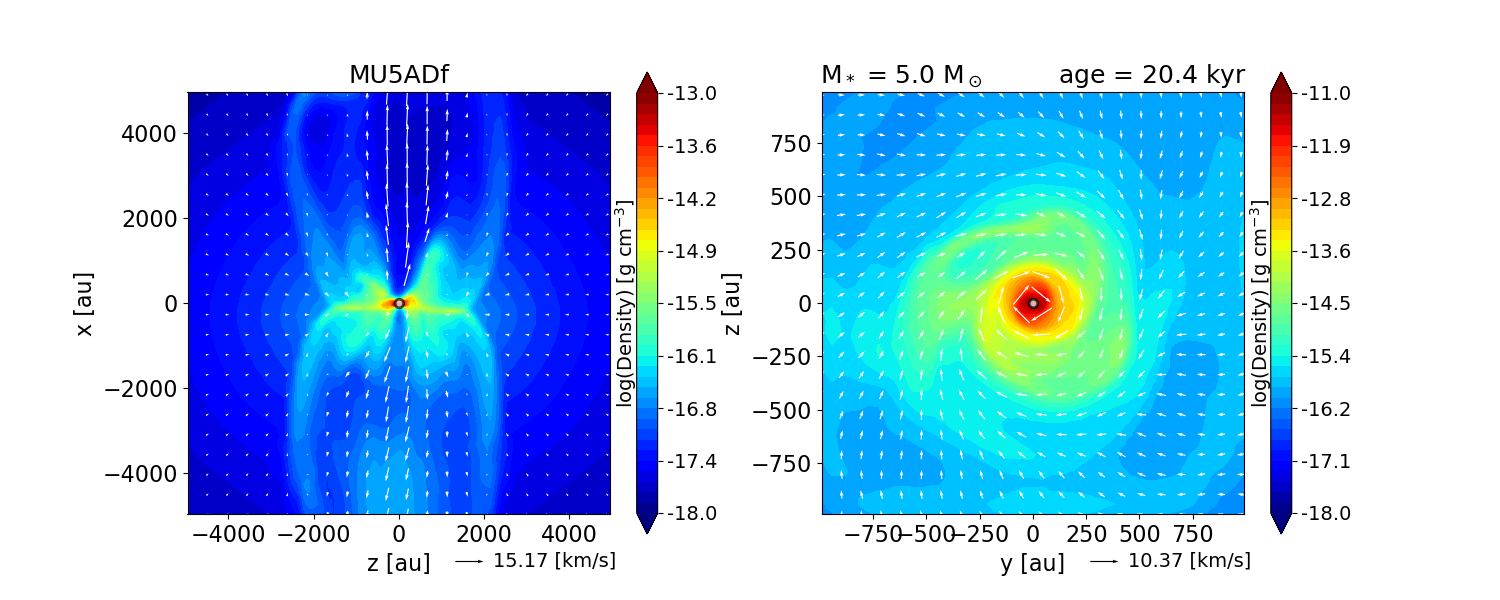}
	\includegraphics[width=0.5\textwidth]{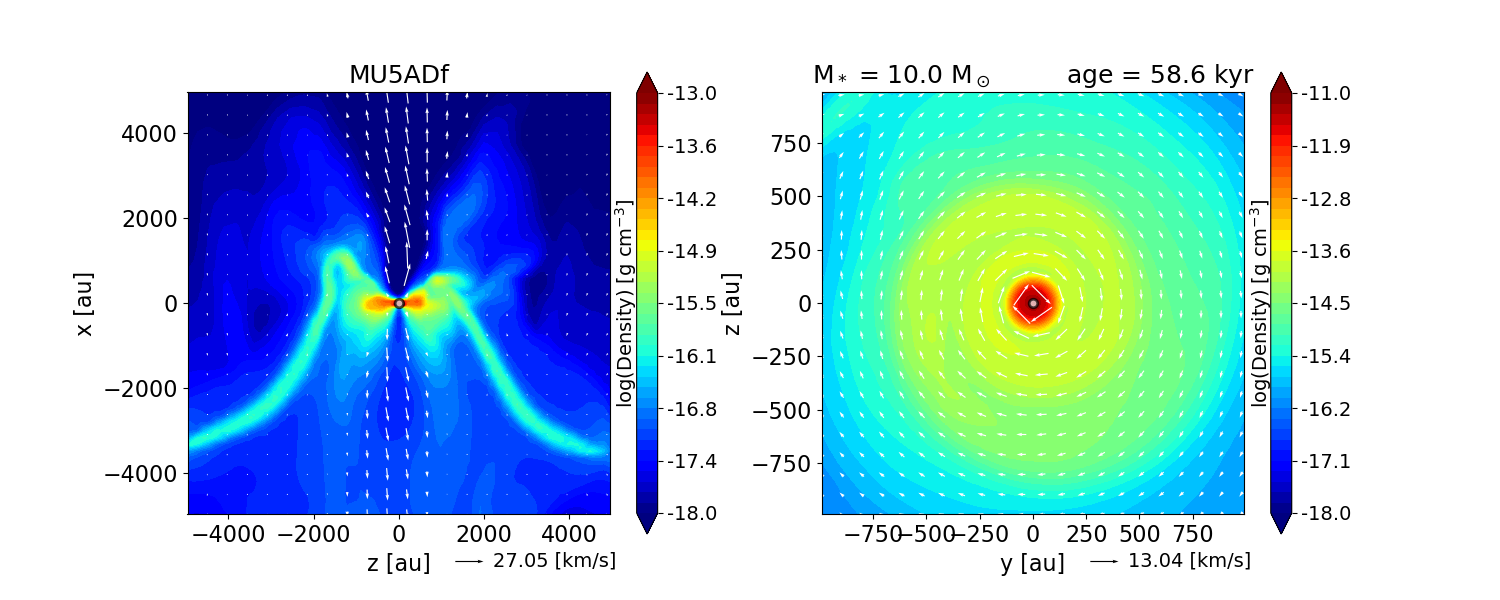}	
	\caption{Density maps in the plane of the disc and perpendicular to it for the models HYDRO (top), MU5I (second line), MU2AD (third line), MU5AD (fourth line), and MU5ADf (bottom). Two different times are represented: $M_\star=5$~\msun~(left) and $M_\star=10$~\msun~(right). The mass $M_\star$ gives the total mass converted into sink particles at each snapshot. The arrows represent the velocity vectors in the plane.
	}
	\label{Fig:densitymaps}
\end{figure*}

Table~\ref{tab:models} reports the time of formation of the first sink particle $t_0$, at which we renormalise the time evolution afterwards. Given our strongly peaked initial density profile and low initial rotation level, all times $t_0$ are very close together, and close to the initial central density free-fall time.

Figure~\ref{Fig:densitymaps} shows edge-on and face-up density maps of the five runs listed in Table~\ref{tab:models} at two different times, when the mass of the central sink particle is 5~\msun~and 10~\msun.

At the beginning of the simulations, the gas collapses toward the centre, until the creation of a sink particle. Around the sink, a rotating disc builds up in the plane perpendicular to the initial solid-body rotation $x$-axis in all models. The largest discs are formed in the HYDRO and MU5ADf runs. The density  in the disc mid-plane is also largest in the MU5I and MU5ADf runs at late time. Outflows are generated in all the magnetised models and propagate along the rotation axis. The largest outflow velocities are found in the runs with ambipolar diffusion where we report maximum velocities of 25-30~km~s$^{-1}$ on 5000 au scales (for a sink mass of $M_\mathrm{sink}=10$~\msun). We observe that none of our models exhibit strong fragmentation, and only the HYDRO model shows secondary sink formation, but these secondary sinks are rapidly merged with the central one.

\begin{figure*}[t]
	\includegraphics[width=1\textwidth]{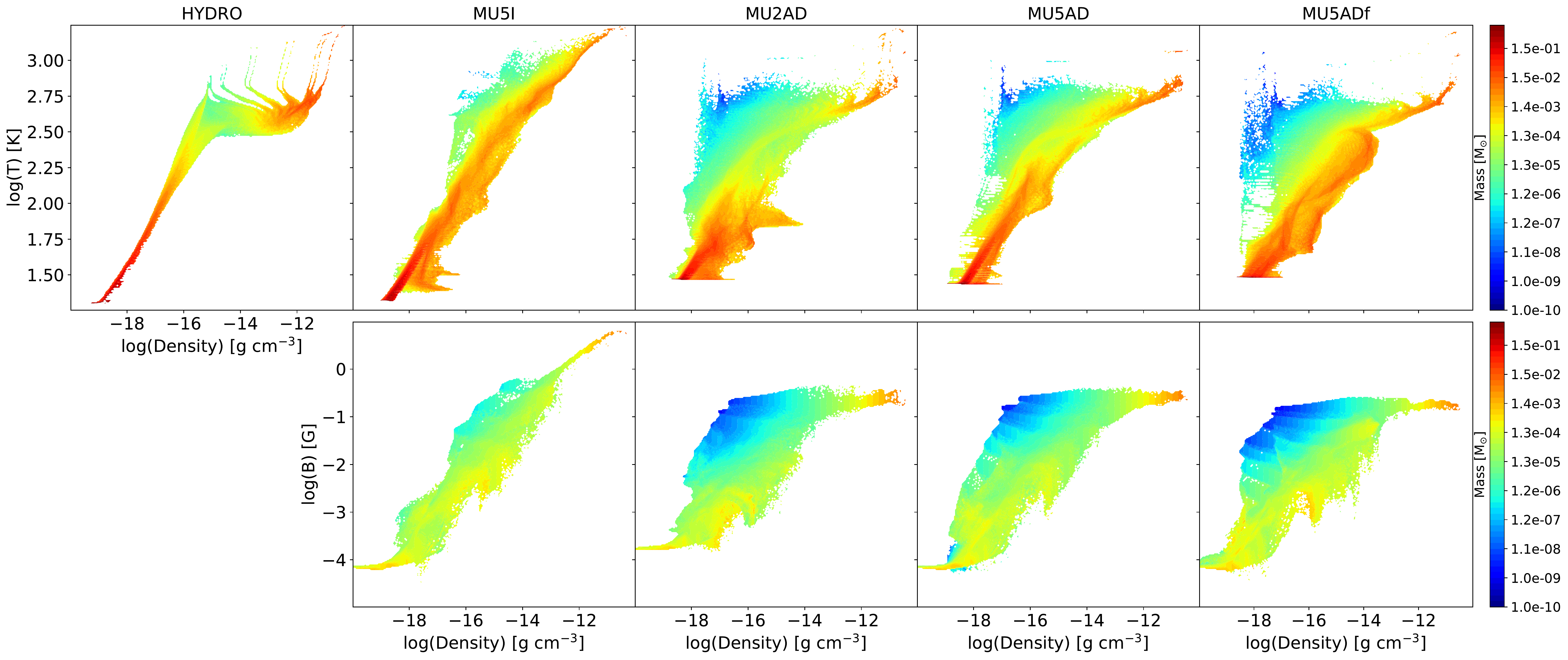}
	\caption{Density-temperature (top panels) and density-magnetic field amplitude (bottom panels) histograms for the models HYDRO, MU5I, MU2AD, MU5AD, and MU5ADf (from left to right) when the central sink mass is 10~\msun. The colour coding indicates the mass.}
	\label{Fig:Trho-Brho}
\end{figure*}

Figure~\ref{Fig:Trho-Brho} displays histograms of density-temperature and density-magnetic field for the five runs when the central sink mass is 10~\msun. All the temperature-density plots look similar, with a central temperature $>1000$~K. One can notice that in the ideal-MHD run, the magnetic field amplitude increases up to value larger than 1~G at high densities as the gas is collapsing as a consequence of the flux-frozen approximation. On the opposite, in all the runs with non-ideal MHD, the magnetic fields stop amplifying at high density. This is due to the strong ambipolar diffusion in the central regions of the collapsing cores. It reaches then a maximum value $<1$~G with a plateau visible in the B-$\rho$ histograms. This result is essentially similar to the one observed in the low-mass star formation regime \citep{masson:16,hennebelle:20} as well as in similar experiments with initial turbulence \citep{mignon:21a}.  Importantly, the efficient ambipolar diffusion enables to redistribute magnetic fields in the low density medium. The region with densities less than $10^{-15}$~g~cm$^{-3}$, i.e. the inner envelop,  exhibits larger magnetization in the models with ambipolar diffusion than in the ideal MHD case. In addition, we observe high temperature ($\simeq 1000$ K) and low density ($\simeq 10^{-18}$ g cm$^{-3}$) regions in the runs with ambipolar diffusion. This corresponds to material within the outflow, close to the protostars ($< 1000$ au), where  the current is strong. As a consequence, the heating by ambipolar diffusion (see Eq. \ref{eq:heating}) is strong in these regions. 

\subsection{Mass evolution \label{sec:mass_evolution}}

\begin{figure}[t]
	\includegraphics[width=0.5\textwidth]{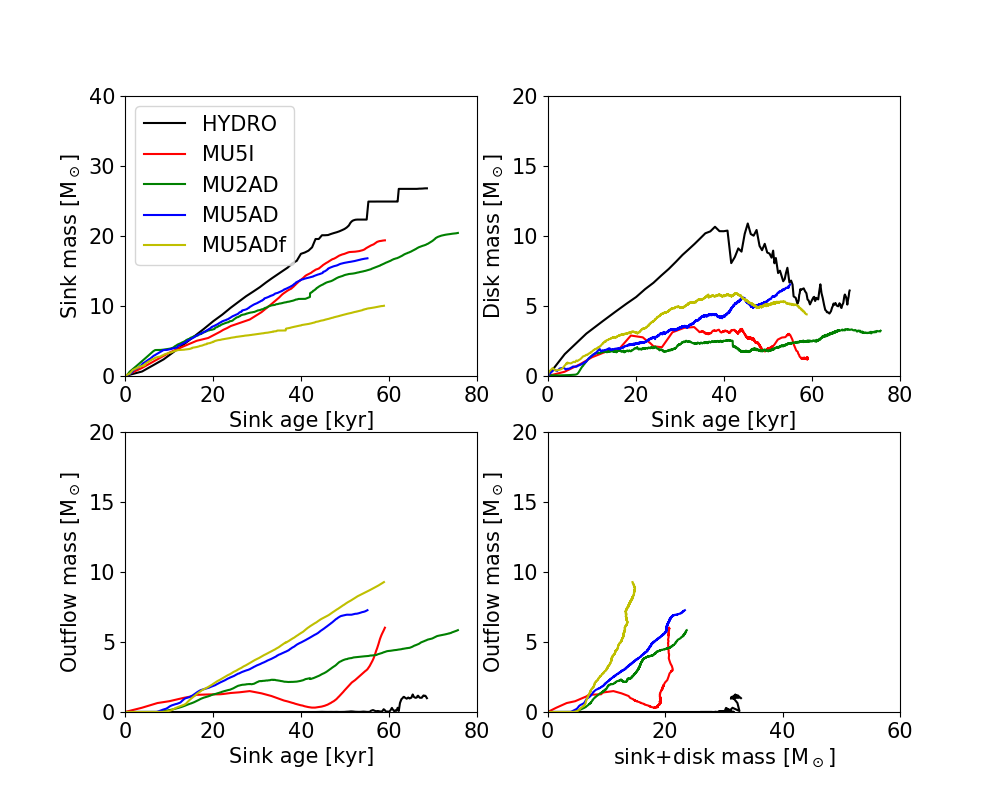}
	\caption{Mass evolution of the sink (top-left), disc (top-right), outflow (bottom-left) as a function of time after sink creation, and of the outflow as a function of the total disc+sink mass (bottom-right). }
	\label{Fig:mass_history}
\end{figure}

Table~\ref{tab:mass} reports the mass of the sink, disc and outflow in the five runs at two different times, when the sink mass equals 5 and 10~\msun. First, the sink mass accretion rate is the smallest for the MU5ADf run, but all sink accretion rates are similar within a factor less than 4. For the outflow, the mass ejection rate is the largest in the resistive runs and negligible in the HYDRO case. Interestingly, the outflow in the ideal MHD case MU5I stalls at about 1  ~\msun.  
In the resistive runs, the mass ejection rate correlates also well with the sink accretion rate, within less than a factor 3. A significant fraction of the mass ($>25\%$ in comparison with the accreted mass onto the sink particle) is thus expelled in the resistive runs. On the opposite, the disc grows faster in mass in the HYDRO case. 

Figure \ref{Fig:mass_history} shows the mass evolution as a function of time of the different components (sink, disc, and outflow) for the HYDRO, MU5I, MU2I, MU5AD, and MU5ADf models.
At first sight, we observe clearly a difference between the HYDRO case and the magnetised ones. The HYDRO run shows the largest sink and disc mass growth and forms an outflow with a negligible mass (we do not report it in Table~\ref{tab:mass}). In addition, the late evolution of the HYDRO central sink mass shows accretion burst events, which are due to mergers with secondary sink particles. The least accreting sink is formed in the MU5ADf . Interestingly, this corresponds to the magnetised model with the largest mass growth of the disc and of the outflow. Then, the resistive runs have outflows of mass larger than 1~\msun, with the largest outflow rate measured in the models with a weak initial magnetic field ($\mu=5$). The outflow in the ideal MHD case MU5I stops being fed and vanishes after about 30 kyr ($M_\mathrm{sink}\simeq 10$~\msun), before it starts again when the sink particle has a mass $>15$~\msun.

From the runs with ambipolar diffusion, we see on the one hand that the time evolution of the sink and disc mass is more dependent on the initial rotation level than on the initial magnetisation. On the other hand, the outflow mass is more dependent on the initial magnetisation. We also do not report any strong universal correlation between the outflow mass and the total mass of the sink and the disc. 

Overall, while the sink mass in all models and outflow mass in the resistive ones only steadily increase as a function of time, the disc mass is not increasing as fast. On the opposite, it shows accretion and ejection phases with time, indicating that the material accreted by the disc is transiting through it to be either ejected or accreted by the sink particle or. 

\begin{figure}[t]
	\includegraphics[width=0.5\textwidth]{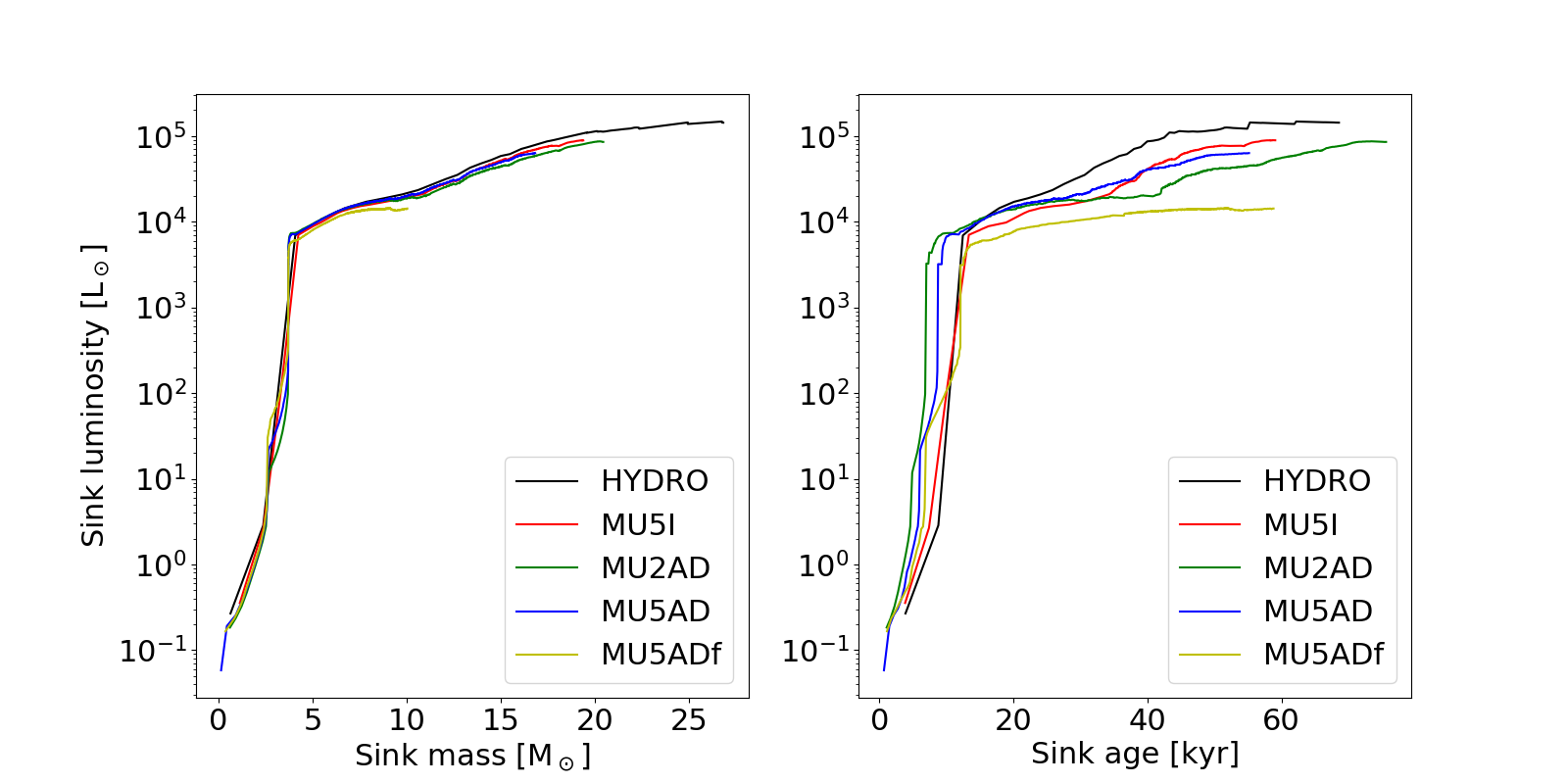}
	\caption{Evolution of the protostellar internal luminosity  as a function of the protostellar mass (left) and of the time after sink formation (right).}
	\label{Fig:lum_history}
\end{figure}

Figure~\ref{Fig:lum_history} shows the evolution of the internal luminosity given by the PMS evolution sub-grid model. First, the luminosity for a given protostellar mass is very similar in all models. This is due to the fact that we do not keep track of the temporal history in the protostellar evolution, so that all models fall on a similar track. Nevertheless, since the accretion rate onto the protostar varies between the models, the time evolution shows variations of about one order of magnitude at an age of 50 kyr, where the HYDRO (resp. MU5ADf) run exhibits the fastest (resp. slowest) increase. We thus expect  the effect of the protostellar radiation to be delayed in the magnetised models. 

We should at this point clarify that the mass evolution of the star-disc-outflow system is barely affected by the formation of secondary sink particles. Only the HYDRO  run forms secondary sink particles. The first secondary sink is formed in the rotation plane (i.e. in the disc)  at a distance of 100~au from the the first sink, which mass is about 16.5~\msun. It reaches a mass of about 0.7~\msun~ before being merged with the central one in less than 1~kyr, when their accretion radii overlap. In total, 10 secondary sinks were also formed in the disc plane at distances of $<500$~au. 9 sink particles quickly merged with the most massive one, migrating inward through the disc in  $<10$~kyr. The maximum mass of the secondary sink particles  before merging is 2.5~\msun.   We note that a small sink particle, of mass $\simeq 0.01$~\msun~gets ejected from the disc. But given the tiny mass of the latter, we did not consider it for further analysis.
Only one secondary sink survives, and forms binary system with the primary one, with separation $\simeq 460$~au and masses of $26.8$ and $6.8$~\msun~at the end of the simulation $t=t_0+69$~kyr. 

\begin{figure*}[t]
	\includegraphics[width=0.5\textwidth]{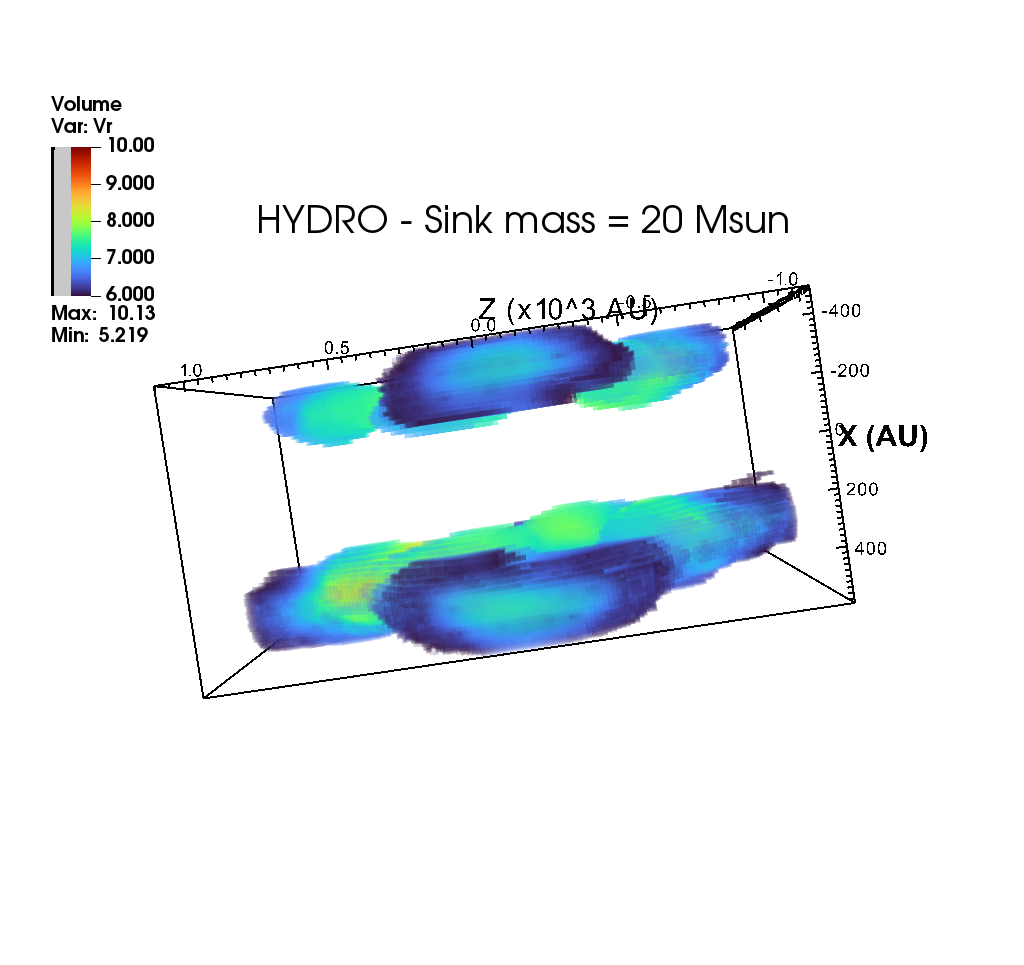}
	\includegraphics[width=0.5\textwidth]{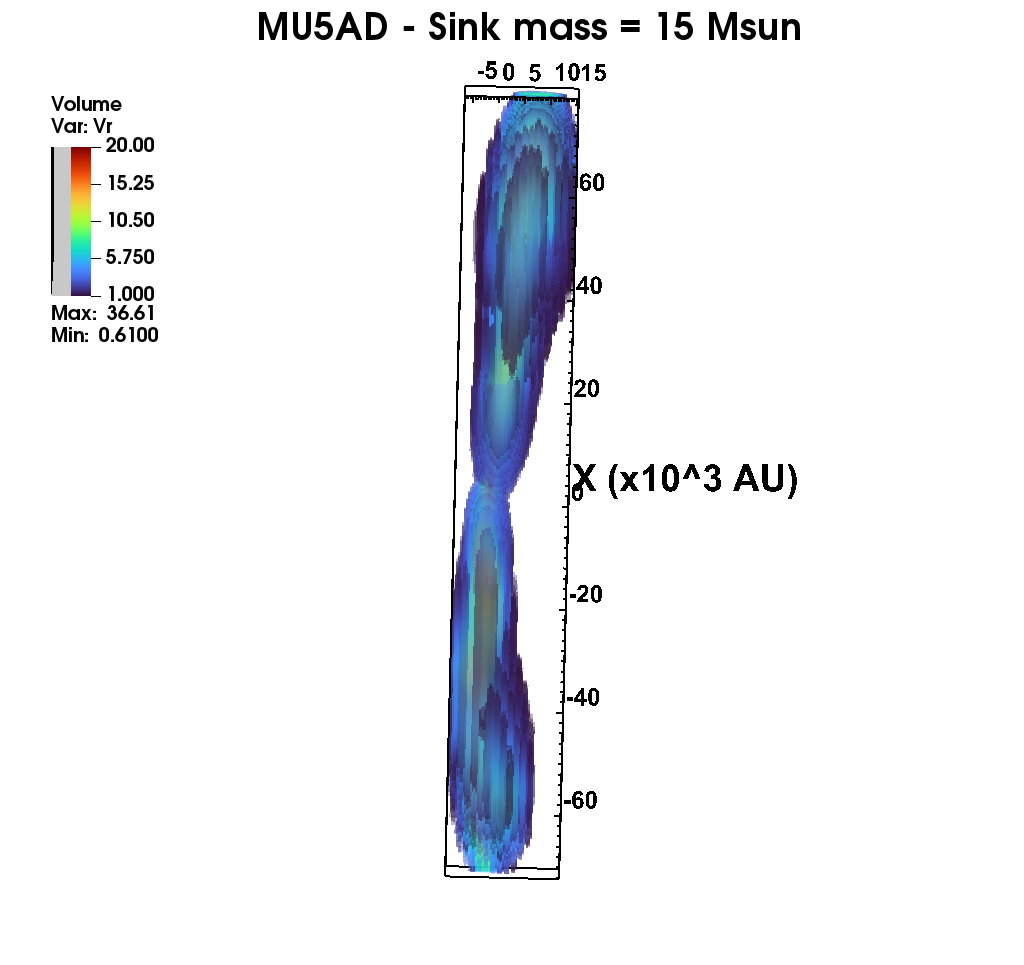}
	\caption{Volume rendering of the outflow in the HYDRO run at the end of the simulation (left) and in the  MU5AD run when the sink mass is 15~\msun~ (right). The colour coding represents the radial velocity and is given in km~s$^{-1}$. In the left panel (HYDRO), the region represents  a box of $\simeq 4000^3$ au, while the vertical extent is of about 75000 au in the right panel (MU5AD run).}
	\label{Fig:morpho1}
\end{figure*}

\subsection{Outflows \label{Sec:outflow}}
In this section, we examine the properties of the outflows, as well as the physical mechanisms responsible of the launching. First, we look at the global morphology of the outflows. Then we study their physical properties.

\subsubsection{Morphology}

All models have outflows. The weakest outflow is found in the HYDRO model as shown in Fig.~\ref{Fig:mass_history}. The left panel of Figure~\ref{Fig:morpho1} shows a volume rendering of the outflowing gas in this case when the sink mass is 20~\msun~and the outflow mass is $\simeq0.01$~\msun. The outflow is made of radiative bubbles,  as observed in previous works by different authors \citep[e.g.][]{krumholz:09,klassen:16}. Indeed, in the HYDRO case,  there is  no other force  than the one provided by the protostellar luminosity that can accelerate the gas against gravitational pull. We note that theses bubble are fragmented by episodic ejections which result from the development of non-axisymmetric features in the rotating region. Since we do not account for the accretion luminosity in our sub-grid radiative feedback module, the episodic events are essentially due to the infalling gas dynamic. Interestingly, the outflow extends more in the radial direction (about 2000~au) than in the vertical one (about 1000 au maximum).   By measuring the different components of the gravitational acceleration,  it turns out that it is  the largest  in the vertical direction since the disk mass contributes  significantly in the gravitational potential . As a consequence, the radiative acceleration, which is essentially isotropic close to the protostar, is the most efficient in the regions of low gravitational acceleration (the radial direction). In addition, the radiation escapes most favourably in the  low optical depth envelop, i.e. just above the disc plane. The anisotropic expansion of the radiative bubbles is thus a consequence of the anisotropic accretion flow on the star-disc system.

The right panel of Figure~\ref{Fig:morpho1} shows the outflow in the fiducial MU5AD case when the sink mass is 15~\msun~and the outflow mass is about 6~\msun. The morphology is completely different than in the HYDRO case with an outflow extending up to 80 000~au in the vertical direction with a maximum velocity more than three times higher. The introduction of the magnetic field changes drastically the outflow morphology. 

\begin{figure*}[t]
\includegraphics[width=1\textwidth]{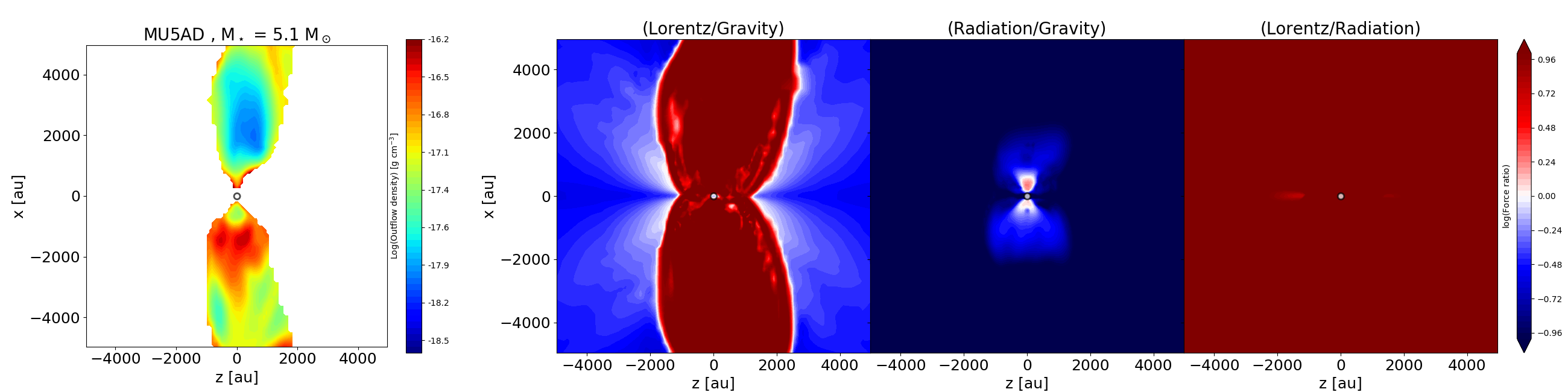}
\includegraphics[width=1\textwidth]{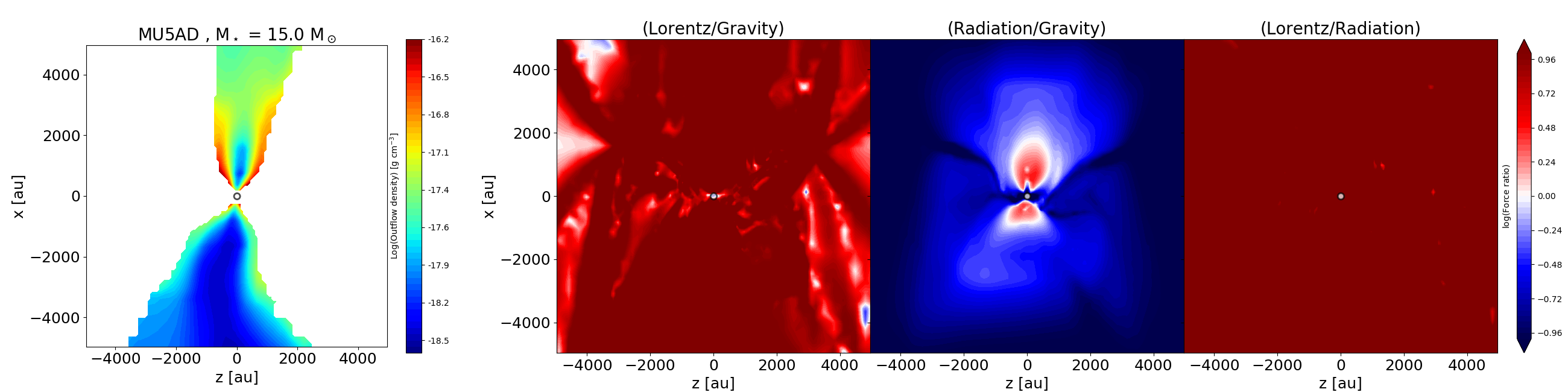}
	\caption{Density and force balance in the outflows for the MU5AD   run when the sink mass is 5~\msun~ (top) and 15~\msun~(bottom).  The left panels show density cuts in the $xz$-plane within the outflow. The three other panels show maps of the ratios between the gravitational, the Lorentz, and the radiative forces.  The colour maps are in logarithmic scale and the force ratio colour map is limited to the range $[-1,1]$ for plot readability, but its value can exceed these values. The bottom row is zoomed in the inner 10 000 au region.}
	\label{Fig:forces}
\end{figure*}

\begin{figure*}[t]
	\includegraphics[width=1\textwidth]{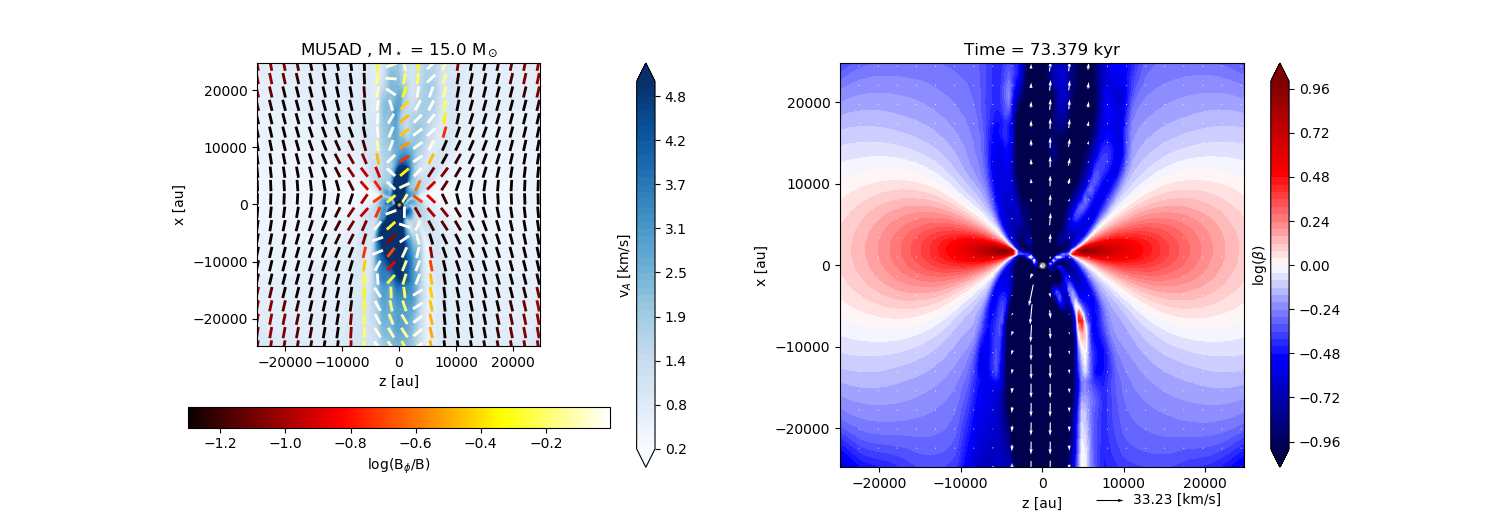}
	\caption{Magnetic fields topology and plasma $\beta=P/P_\mathrm{mag}$ in the $xz$-plane  in the MU5AD run, when  the sink mass is 15~\msun. The left panels show the amplitude of the Alfvén velocity (blue colours) and the segments show the magnetic fields direction in the $xz$-plane. The colour coding represents the ratio between the toroidal component of the magnetic fields over the total magnetic fields. Yellow to white segments exhibits strong toroidal fields with $B_\phi/B>0.3$. The right panels show maps of the plasma $\beta=P/P_\mathrm{mag}$ with velocity vectors overplotted. The red regions are dominated by the thermal pressure while the blue ones are dominated by the magnetic pressure. }
	\label{Fig:Bphi}
\end{figure*}

\subsubsection{Outflow origin}

The morphology of the outflows we report above suggests that different mechanisms are at play. In the HYDRO case, as previously mentioned, there is no doubt that the outflow material is accelerated by the radiative force. We note that the outflow appears about 40 kyr after the sink formation, when the sink mass is almost 20~\msun.  The outflow hardly develops and remains with a limited extent of about 1000 au in the vertical direction.

 Figure~\ref{Fig:forces} shows maps of the density and of the force ratio in the outflows for the MU5AD  run when the  sink mass is 5 ~\msun~ (top) and 15~\msun~ (bottom). The outflow remains similar in shape, although it gets broader with time. In the entire outflow region, the Lorentz force dominates by more than one order of magnitude the gravitational one. Closer to the sink particle up to a distance of 2 000 au, the radiative force also dominates over the gravitational one, the extent of  this region increasing with the stellar mass. The radiative and Lorentz forces thus both contribute to the outflow launching. When comparing the Lorentz and radiative forces (right column), the Lorentz force clearly dominates  everywhere the force balance and is thus the main contributor to the acceleration of the outflowing gas. The outflow in the MU5AD case is thus of magnetic origin.
	We put nevertheless a note of caution, since the grey approximation underestimates the magnitude of the radiative force, the latter being larger if one accounts for frequency dependent irradiation \citep[e.g.][]{kuiper:10,mignon:20}. We discuss this limitation in Sect.~\ref{Sec:discussion_limits}.

At a stellar mass of 15~\msun, the radiative acceleration is greater than the gravitational one over a larger extent in the MU5AD case than in the HYDRO case. This is due to the low-density cavity created by the magnetic outflow in the vertical direction, which facilitates the radiation escape in a low optical depth region. In the HYDRO case, the density is too high because of the collapsing envelope and the radiation cannot accelerate the gas sufficiently to escape.

\begin{figure*}[ht]
	\includegraphics[width=\textwidth]{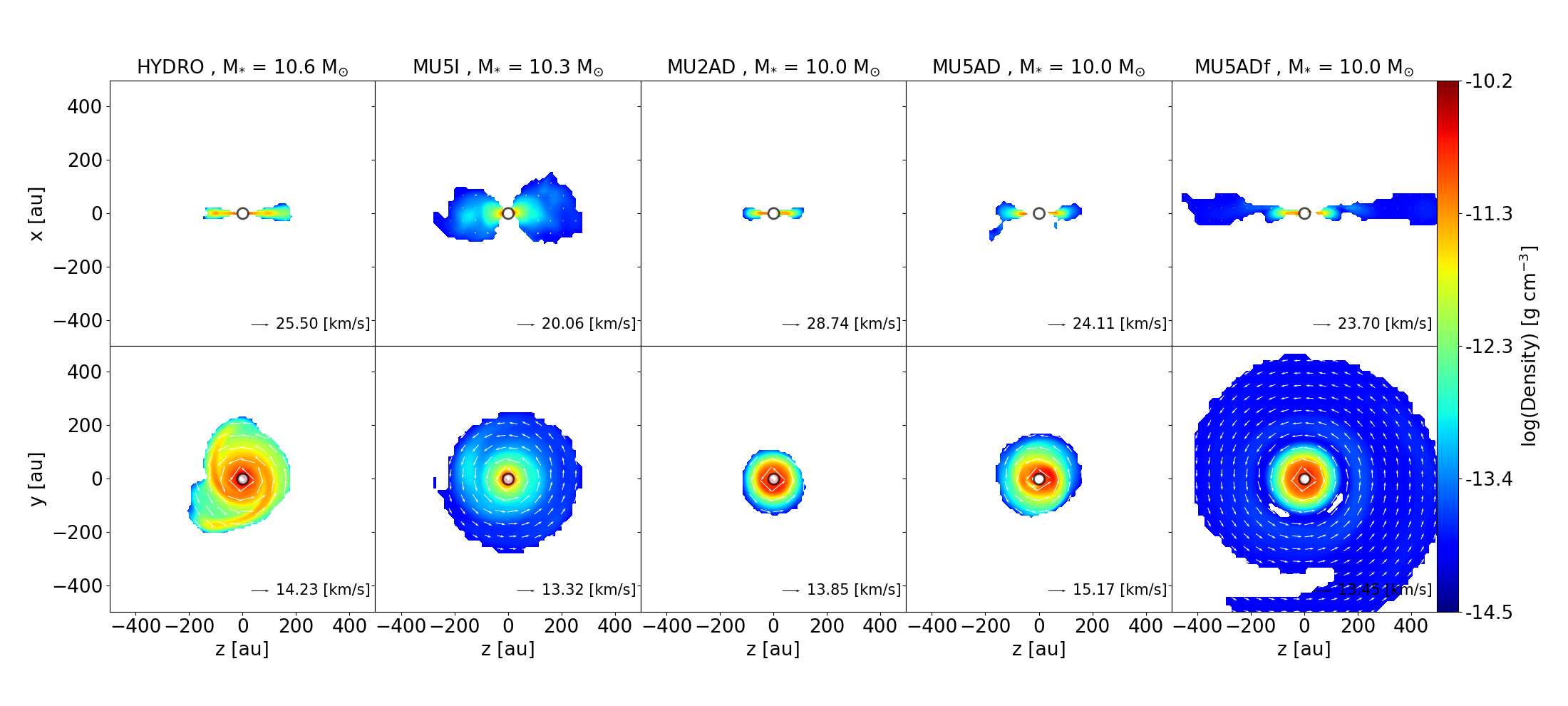}
	\caption{Density maps perpendicular to the disc plane (top) and in the disc plane (bottom) when the sink mass is $\approx 10$ \msun~for the runs HYDRO, MU5I, MU2AD, MU5AD, and MU5ADf, from left to right. The vectors indicate the gas velocity. Only the material of the disc is plotted.}
	\label{Fig:disc}
\end{figure*}

Figure~\ref{Fig:Bphi} illustrates the magnetically-driven origin of the outflow in the MU5AD run when the sink mass is 15~\msun. Alfv\'en velocity and magnetic fields topology are shown in the left panels, and plasma $\beta=P/P_\mathrm{mag}$ and velocity in the right panels.
The toroidal component of the magnetic field clearly dominates in the outflow region, which is characteristic of self-collimated magnetically driven outflow. Correspondingly, the outflow region is dominated by the magnetic pressure with $\beta<1$. We observe the same features in the other runs with ambipolar diffusion  (MU2AD, and MU5ADf) throughout their evolution after the launching of the outflow. A detailed analysis on the different MHD mechanisms at play in outflow launching  (magnetic pressure versus magneto-centrifugal acceleration)  goes beyond  the scope of the present study. We invite the reader to refer to \cite{mignon:21b} for a dedicated study on the outflow acceleration in similar models including ambipolar diffusion and turbulence.

In the ideal MHD  MU5I run, we already noted in Sect.~\ref{sec:mass_evolution} that the outflow almost disappears  when the sink mass is 10~\msun. The outflow mass starts decreasing at mass $\simeq 4$~\msun. Before this, the outflow is very similar to the one in the ambipolar diffusion case, except that the magnetic fields amplitude is larger in the central region because of the magnetic field amplification resulting from the ideal MHD approximation. We also note that in this case, the outflow is launched at the same time as the sink particle forms and as a consequence, the outflow is broader and has a larger extent compared to the ambipolar diffusion runs. At a mass $>4$~\msun, the PMS luminosity strongly increases ($L>10^3 L_\odot$) as can be noted from Fig.~\ref{Fig:lum_history}, which  causes a brutal increase of the thermal pressure in the sink particle vicinity. This large thermal pressure variation, combined with the large magnetic pressure resulting form the ideal MHD approximation, results in a destabilization of the central region and in a kick of the central sink particle as observed for instance in numerical works reporting interchange instability \citep[e.g.][]{zhao:11}. As a consequence,  the magnetic structure of the outflow base gets disrupted and the MHD driving of the outflow stops. This behaviour is not observed in the runs with ambipolar diffusion, since the magnetic fields amplification is prevented in the central regions (see Fig. \ref{Fig:Trho-Brho}). In addition, the material at the outflow base gets preheated by the friction provided by the strong ambipolar diffusion heating. We note that the outflow restarts in the MU5I run when the central sink mass gets larger than 15~\msun. As can be seen from Fig.~\ref{Fig:mass_history}, the outflow mass loading is very efficient at this stage in the MU5I model. Indeed, both, the radiative and the Lorentz force accelerations are much larger than at earlier times and participate to the launching of the outflow.

To summarise, the outflows formation and evolution completely change when magnetic fields are introduced in the models. We note  that ambipolar diffusion stabilises the outflow driving at early stages when the sink mass is less than 20~\msun. A more detailed analysis of the outflow launching and structure is provided in the follow-up study \cite{mignon:21b}.

\subsection{Properties of the disc \label{Sec:disc}}

\subsubsection{Disc size}

\begin{figure}[t]
	\includegraphics[width=0.5\textwidth]{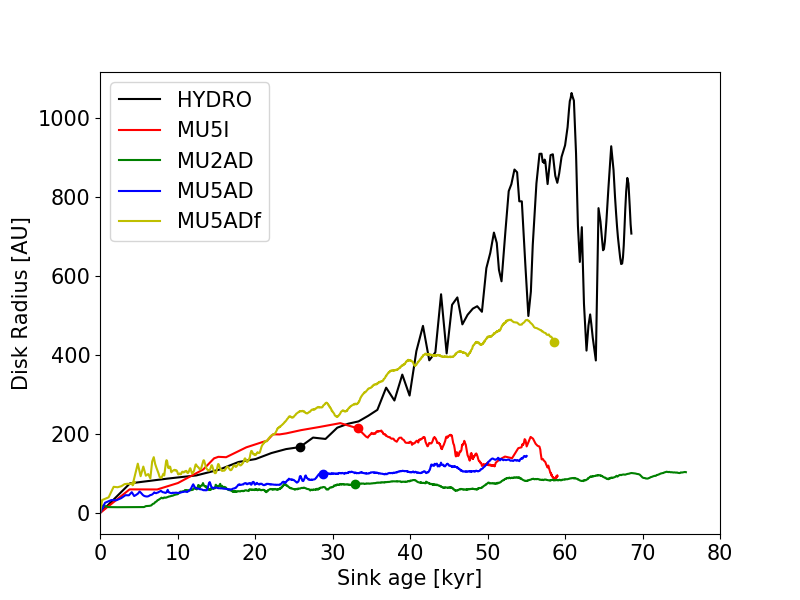}
	\caption{Time evolution of the disc radius for the HYDRO, MU5I, MU2AD, MU5AD, and MU5ADf  runs. The symbols on the curves show the time at which the sink mass is 10~\msun.}
	\label{Fig:radius_disc}
\end{figure}

\begin{figure}[t]
	\includegraphics[width=0.5\textwidth]{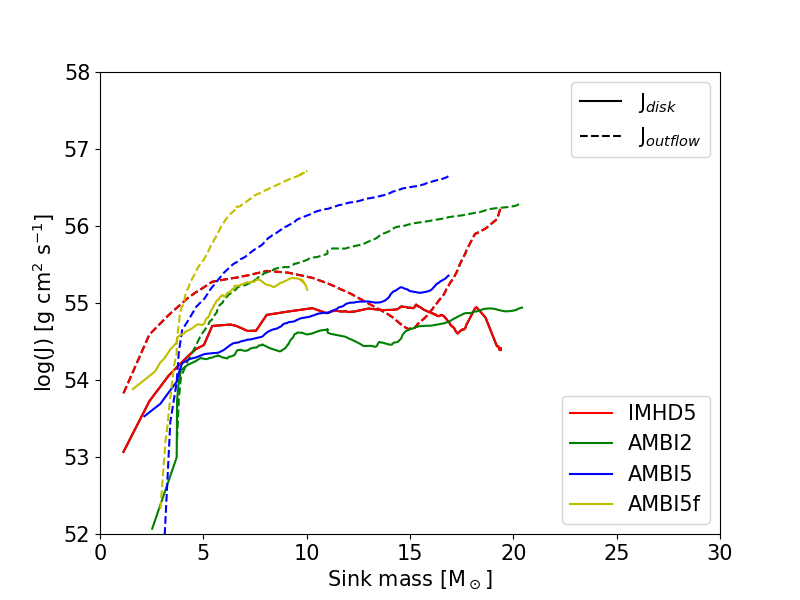}
	\caption{Evolution of the angular momentum contained in the disc (solid line) and in the outflow (dashed) in the magnetised models  MU5I, MU2AD, MU5AD, and MU5ADf . The symbols on the curves show the time at which the sink mass is 10~\msun.}
	\label{Fig:angmom}
\end{figure}

Figure~\ref{Fig:disc} shows density maps of the disc material  in the equatorial plane and in the perpendicular direction in the HYDRO, MU5I, MU2AD, MU5AD, and MU5ADf runs when the sink mass is 10~\msun. The disc formed in the HYDRO run  has a radius of about 170 au and a mass of 7.2~\msun. The disc develops spiral arms which are a sign of instability, but  the disc does not fragment. In the MU5I run, the disc is more extended with a radius of about 220~au and a mass of 3.5~\msun. The  disc is also puffier with a vertical height of about 100~au. The disc does not develop strong spiral arms in this case, and does not experience fragmentation. 
The discs formed in the MU5AD and MU2AD runs are very similar. First, they are much smaller with a radius less than 100~au (98 and 73~au respectively) and less massive (3.3 and 2.4~\msun~resp.).  Interestingly, the disc radius is smaller in the MU5AD run than in the MU5I one, whereas we would expect that the ideal MHD framework provides a more efficient magnetic braking as it is widely observed in low-mass star formation (see below). 
Last, the disc in the MU5ADf case is the largest one, but its material remains at a low density compared to the HYDRO case.  It exhibits a spiral arm at large radii, but it does not fragment. 

Figure~\ref{Fig:radius_disc} shows the time evolution of the disc radius for the same models as in Fig.~\ref{Fig:disc}. The MU5AD and MU2AD runs show a very similar time evolution for the disc size, increasing slowly with time. The disc radius in the HYDRO case increases the most, to reach about 800~au at the end of the simulation. The disc exhibits periods of expansion and contraction of a few kyrs while globally expanding. This succession of expansion/contraction events is due to the apparition of prominent spiral arms (see below). In the MU5ADf case, the disc  increases  rapidly and is the largest one when the sink mass is 10~\msun. As previously noted, even if the disc is large in radius, its density is low at large radii and its mass is comparable to the one in run MU5AD. The disc in the MU5I run stops increasing at about 200~au at time 30~kyr. It shrinks after 30~kyr, which roughly corresponds to the time when the destabilization of the central part due to high magnetic and thermal pressure is the strongest (the MHD outflow stops being driven). 

We now focus on a puzzling observation: the disc radius in the MU5I model is larger than the one in the MU5AD whereas it is expected that ideal MHD enhanced magnetic braking and prevents disc formation \citep[see][for the low-mass star case]{hennebelle:08}. Firstly, we stress that given our initial conditions, the magnetisation corresponds to a weak field case ($\mu_\mathrm{c}=50$ for the MU5 runs). It is thus expected with such low magnetisation that discs form even in the ideal MHD limit \citep{commercon:10,seifried:12}. Initially, for sink mass $<5$~\msun, we not the that the discs are comparable in size whatever MHD framework. Then, the disc is larger in the MU5I case is only a factor less than 2 larger than the MU5AD case with a maximum difference around 30 kyr. 
Secondly, we compare the efficiency of the magnetic braking in the envelope for the four magnetised models and we measured that it is 1) the main angular momentum transport mechanism in the envelope 2) of the same order of magnitude in all models (not shown here for readability). The surrounding of the disk is indeed highly ionised so that magnetic braking is efficient in the envelope even in the resistive case \cite[see][for the low mass case]{lee:21}. The magnetic fields amplitude measured in fig.~\ref{Fig:Trho-Brho} indeed shows that there are of the same order in the envelope close to the disc (density close to $10^{-15}$ g cm$^{-3}$). Instead, we find that the amount of angular momentum carried out by the outflow material varies a lot if ambipolar diffusion is taken into account. Figure~\ref{Fig:angmom} shows the evolution of the angular momentum contained in the disc and in the outflow as a function of the central sink particle mass. The MU5ADf  exhibits the highest angular momentum since it is the model with initially a higher rotation level. The disc angular momentum in the three other models is similar within less than one order of magnitude. On the opposite, the outflow carries angular momentum very differently depending on the MHD approximation. In the resistive case, the angular momentum in the outflow increases continuously with the central mass.  In the MU5I case, the amount of angular momentum follows what we previously observed for the outflow driving. It first increases as soon as the outflow gets launched, but it then decreases when the sink mass exceeds $\simeq 8$~\msun. It starts increasing again at central masses $>15$~\msun~ when the outflow gets driven again. In addition, if we focus on the ratio between the outflow and the disc angular momentum, we see that it is smallest for the MU5I. As the outflow is driven mainly on disc scales, it indicates that the outflow is less efficient extracting angular momentum from the disc in the MU5I case. Since the magnetic braking on the envelope is similar in all models, the angular momentum left into the disc is largest in the MU5I case and this explains why the disc grows faster in this case. Last, we also note that the disc is more diluted in the MU5I case while the mass is similar (for time less than 40 kyr). As a consequence, the disc gets larger.

\subsubsection{Disc stability}
\begin{figure*}[t]
	\includegraphics[width=\textwidth]{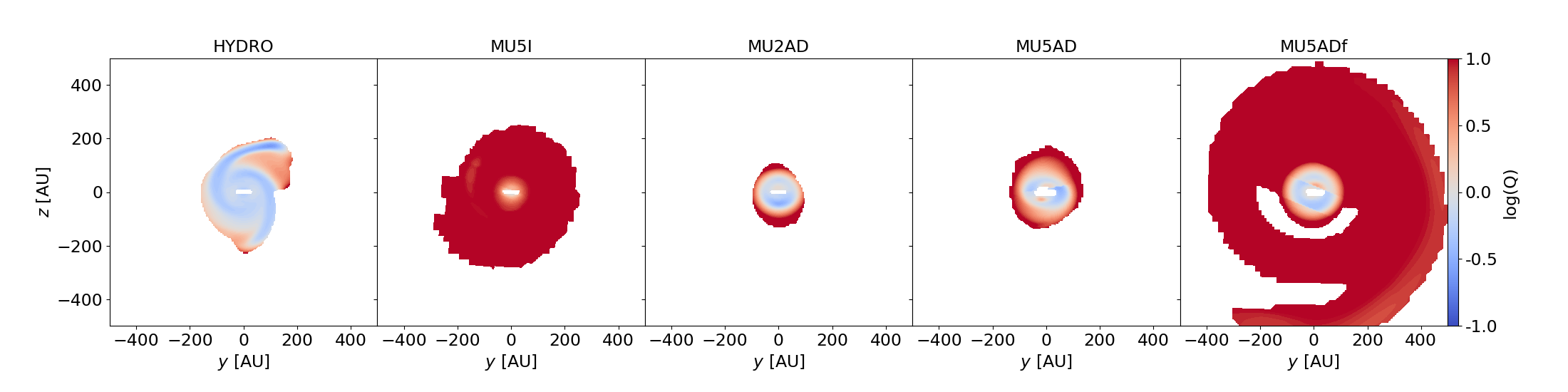}
	\caption{Maps of the Toomre parameter $Q$ in the rotation plane when the sink mass is $\approx 10$ \msun~for the runs HYDRO, MU5I, MU2AD, MU5AD, and MU5ADf, from left to right. Only the disc material is used to compute the Toomre parameter as explained in the main text. The colour map indicates the logarithm of $Q$, with blue (resp. red) colour indicating $Q<1$ (resp. $Q>1$) regions.}
	\label{Fig:Toomre}
\end{figure*}

We already noted that the discs experience gravitational instability and spiral arms formation. A useful parameter to quantify the stability of the rotationally supported disc is the Toomre parameter \citep{toomre:64}, defined as
\begin{equation}
Q=\frac{C_\mathrm{s}\kappa}{\pi G \Sigma},
\end{equation}
where $C_\mathrm{s}$ is the gas sound speed, $\Sigma$ is the disc surface density (in g~cm$^{-2}$), and $\kappa$ is the epicyclic frequency. For Keplerian discs, the epicyclic frequency $\kappa$ is equal to the angular velocity $\Omega$, defined as  $\Omega = v_\mathrm{Kepl}/R_\mathrm{disc}$, with $v_\mathrm{Kepl}= (G M_\mathrm{disc}/R_\mathrm{disc})^{0.5}$.

We note that we show in the next paragraphs and in Fig.~\ref{Fig:profile_Bv} that the Keplerian rotation  is a good approximation in the disc. 

We compute the Toomre parameter locally in the rotation plane as $Q(y,z)$ by averaging (mass-weighted) the sound speed and the angular velocity over the disc height as follows (assuming the rotation axis is the $x$-axis)
\begin{equation}
\bar{\Omega}(y,z)=\frac{1}{\Sigma(y,z)}\int \frac{v_\theta(x,y,z)}{r}\rho(x,y,z) dx.
\end{equation}

Figure~\ref{Fig:Toomre} shows maps of the Toomre parameter $Q$ in the rotation plane for the disc material when the central sink particle mass is 10~\msun. The most unstable disc is found in the HYDRO case, with almost all the disc material exhibiting $Q<1$. On the opposite, the disc in the MU5I is stable everywhere. Indeed, the extra support provided by magnetic fields makes the disc puffy and the density is thus lower as seen on Fig.~\ref{Fig:disc}. In the resistive runs, we see that the disc is stable in the outer part while it is unstable in the inner part. The unstable regions corresponds to the one where the temperature, as well as the plasma beta are the highest. Interestingly, we find that the size of the unstable region is very similar between all the resistive runs and is about 100~au.  We put a note of caution on the size of the unstable region in the disc since its absolute value can be affected by our choice of neglecting the accretion luminosity and of the isotropic FLD irradiation \citep[e.g.][]{mignon:20}.

We note that we do not take into account the Alfv\'en velocity in the Toomre parameter estimate as can be found in \cite{kim:01} or \cite{vaytet:18}. The disc in the cases with ambipolar diffusion or without magnetic fields are dominated by the thermal pressure, so that we can neglect the magnetic support. On the opposite, we may nevertheless underestimate the Toomre parameter in the MU5I model where the Alfv\'en speed is found to be slightly larger than the sound speed (see Fig.~\ref{Fig:profile_Bv}). This would however not change our results since the disc in the MU5I is already the most unstable.

\subsubsection{Properties of the magnetised discs}

\begin{figure}[t]
	\includegraphics[width=0.5\textwidth]{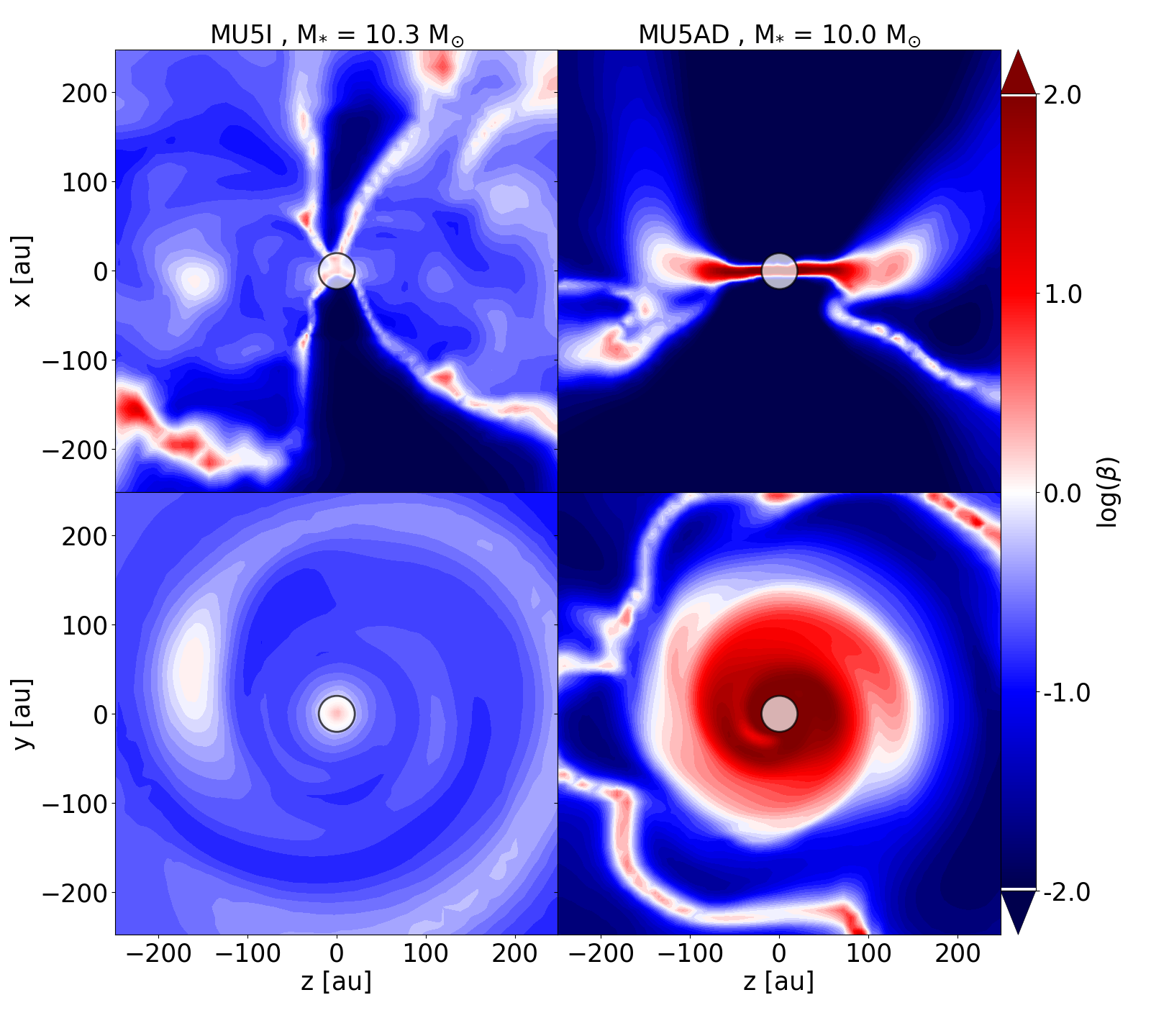}
	\caption{Plasma $\beta$ for the MU5I (left) and MU5AD (right) runs at the same times as in Fig.~\ref{Fig:disc}. The top (bottom) row shows edge-on (face on) cut. Scales of $\beta$ are logarithmic.}
	\label{Fig:plasmabeta}
\end{figure}

In this section, we focus on the properties of the disc formed in the magnetised models. In particular, we discuss the fundamental differences between the ideal MHD and the resistive runs. Figure~\ref{Fig:plasmabeta} shows the plasma $\beta$ in the disc and close envelope when the sink mass is 10~\msun~for the runs MU5I and MU5AD. In the ideal MHD run, all the disc region exhibits $\beta<1$, i.e it is dominated by the magnetic pressure.  On the opposite, the disc in the MU5AD run shows $\beta\gg 1$ in the inner disc region, i.e. it is dominated by the thermal pressure by more than two orders of magnitude. The magnetic pressure is thus negligible when the disc evolves at early times if  ambipolar diffusion is active. We note that similar results are found in the low-mass star formation regime \citep{masson:16,lam:19,hennebelle:20}, with discs in the resistive models dominated by the thermal pressure. These results have been further confirmed in \cite{mignon:21a} where we run similar models with ambipolar diffusion, but accounting for initial turbulence and using a better irradiation scheme. 

\begin{figure*}[t]
	\includegraphics[width=1\textwidth]{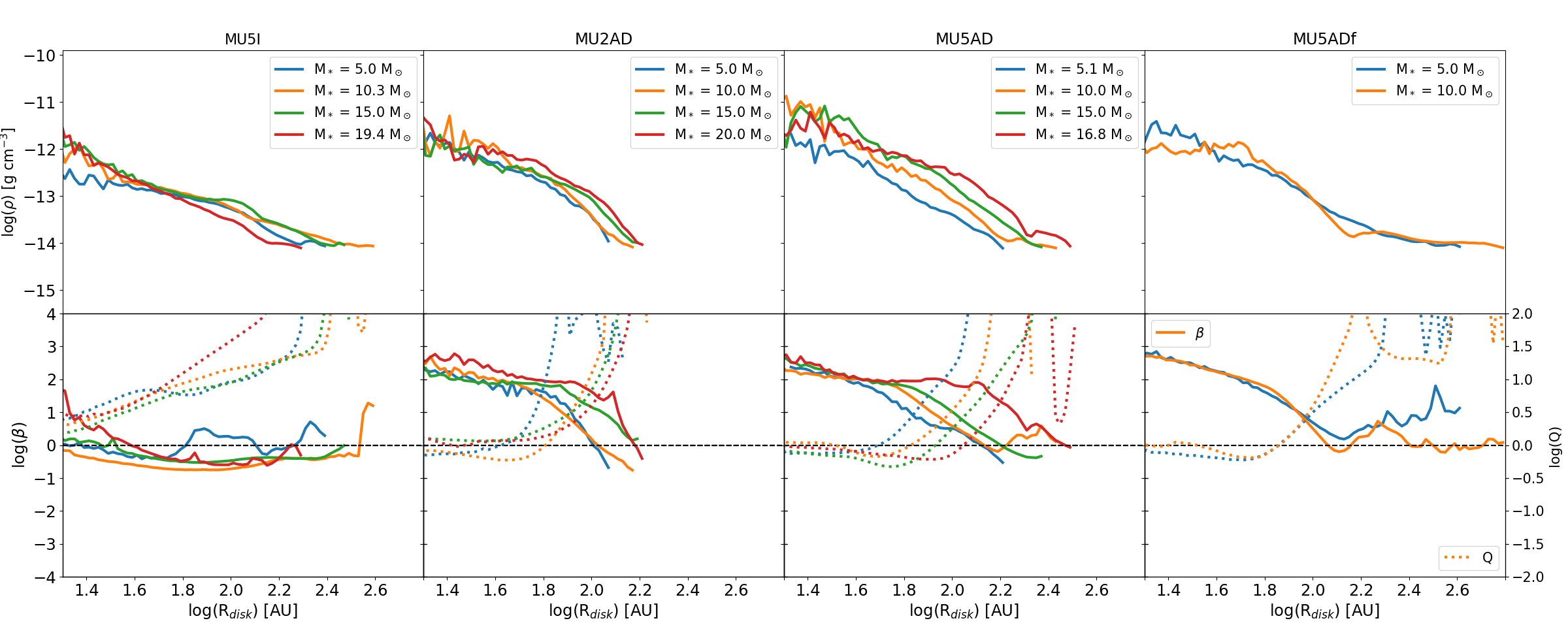}
	\caption{Mean radial profiles of the density (top), Toomre parameter (bottom, dotted line) and plasma $\beta$ (bottom, solid line) at different times for the MU5I, MU2AD, MU5AD, and MU5ADf runs, from left to right. }
	\label{Fig:profile_rhobeta}
\end{figure*}

  Figure~\ref{Fig:profile_rhobeta} shows the mean density, the Toomre parameter  and the density-averaged plasma $\beta$ radial profiles within the disc. We take the arithmetic mean weighted by the density as a function of the radius, averaging over azimuths and height. The density profiles are broadly similar in the models with ambipolar diffusion, with a trend of increasing density with time in the models with slow rotation (MU2AD and MU5AD). The density profile is shallower  in the MU5ADf run, the disc being more massive and also more extended. In the MU5I run, the disc inner density is lower than in the resistive runs, as a consequence of the extra support provided by the toroidal magnetic pressure which lift the gas in the vertical direction. We also note that the column density in the vertical direction is also lowest in the MU5I at radii $< 100$~au. The corresponding plasma $\beta$ shows essentially opposite features between ideal and resistive MHD. In the resistive runs, the inner parts of the disc have $\beta>1$ as already mentioned. Interestingly, the largest $\beta$ is found in the fast rotation model MU5ADf, probably a consequence of the smaller accretion rate of the sink and the disc  which allows more time for ambipolar diffusion to operate before reaching 10~\msun~(see Table~\ref{tab:mass}). In the MU5I case, the disc is dominated by magnetic pressure for sink mass $<20$~\msun. The Toomre parameter shows also opposite behaviours It is larger than 1 at all radii in the MU5I run. In the resistive runs,  it is less than unity in the inner region of the disc and larger than unity in the external parts. The size of the unstable region tend to increase with time in particular in the MU5AD case.

\begin{figure*}[t]
	\includegraphics[width=1\textwidth]{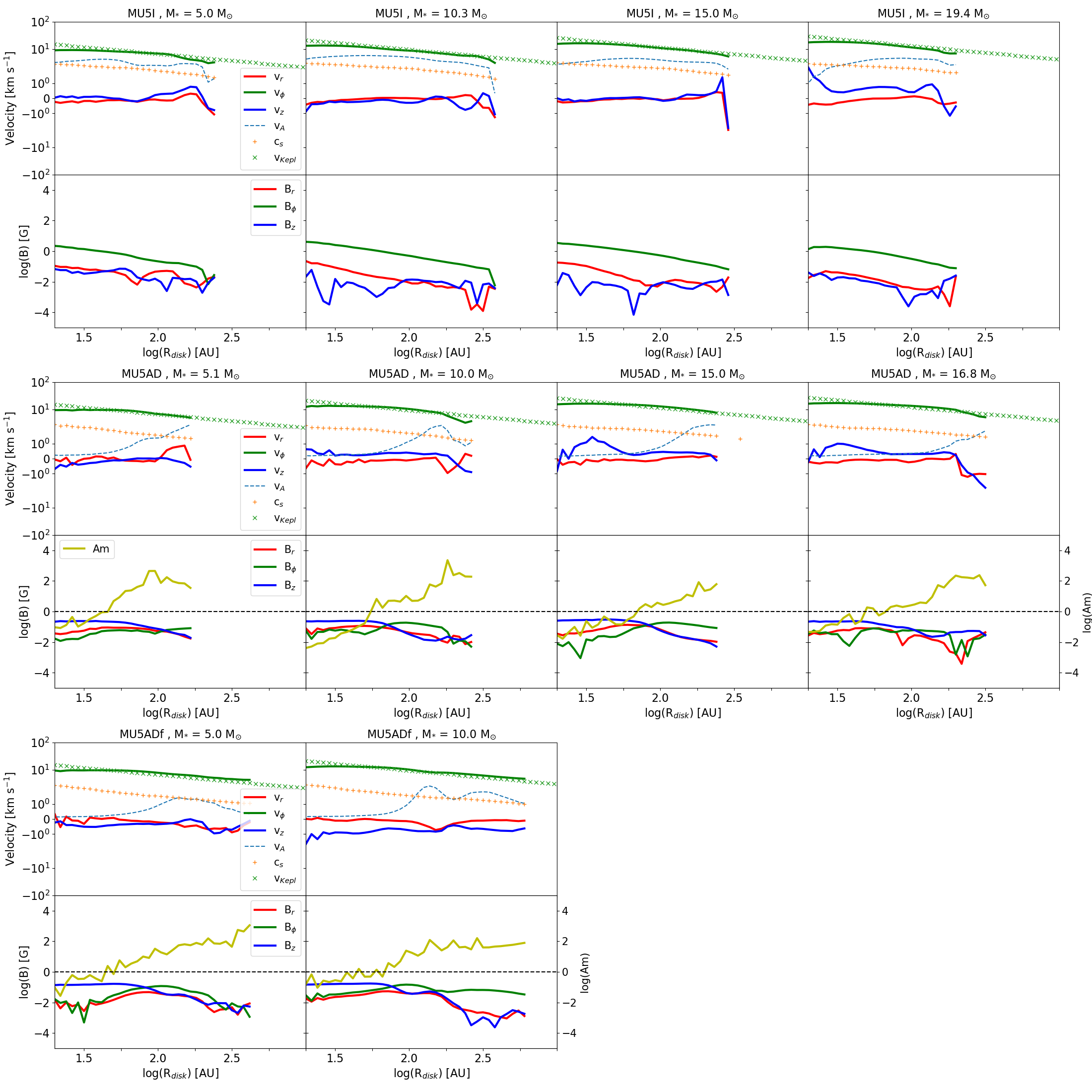}
	\caption{Mean radial profiles of the velocities and magnetic field components for the MU5I (two first rows), MU5AD (rows 3 and 4), and MU5ADf (rows 5 and 6).  The time (and sink mass) increases from left to right. In the velocity plots, we represent the radial velocity $V_\mathrm{r}$ (thick red line), azimuthal velocity $V_\mathrm{\phi}$ (thick green), vertical velocity $V_\mathrm{z}$  (thick blue), Alfv\'en velocity $V_\mathrm{A}$ (dashed blue), sound speed $C_\mathrm{s}$ (green cross), and Keplerian velocity $V_\mathrm{K}=(GM_\mathrm{sink}/R)^{0.5}$ (red cross). In the magnetic components panels, we show the logarithm of the absolute value of the radial component $B_\mathrm{r}$ (thick red), the toroidal component $B_\mathrm{\phi}$ (thick green), vertical component $B_\mathrm{z}$  (thick blue), as well as the ambipolar diffusion Elsasser number $\mathrm{Am}$.}
	\label{Fig:profile_Bv}
\end{figure*}

We now investigate the properties of the velocity and magnetic fields in the disc. To quantify the relative importance of ambipolar diffusion over other dynamical effects, we define the Elsasser number for ambipolar diffusion $\mathrm{Am}$ as the ratio between the rotation time $\Omega^{-1}$ and the ion-neutral collision time $t_\mathrm{in}=\eta_\mathrm{AD}c^2/(4\pi v_\mathrm{A}^2)$
\begin{equation}
\mathrm{Am}=\frac{4\pi v_\mathrm{A}^2}{c^2 \eta_\mathrm{AD}\Omega},
\end{equation}
where $\Omega$ is the Keplerian rotation frequency.

Figure~\ref{Fig:profile_Bv} represents the radial profiles within the disc of the mean velocities and magnetic field components for the MU5I, MU5AD, and MU5ADf runs. In addition, we show the mean Elsasser number profile in the magnetic components plots of the resistive runs (right axes). The averaging is done the same way as in Fig.~\ref{Fig:profile_rhobeta}. In all the models, the velocity is dominated by the azimuthal component, which matches very well the  Keplerian velocity throughout the disc. The radial and vertical components are negligible. As expected,  the $\beta>1$ (resp. $<1$) regions exhibits  Alfv\'en velocities smaller (resp. larger) than the sound speed. The magnetic field topology in the MU5I is clearly dominated by the toroidal component at all masses. On the opposite, the vertical component dominates in the inner part of the disc in the resistive runs. This result is in agreement with what we find in previous work in the context of low-mass star \citep{masson:16,hennebelle:16,hennebelle:20} in the regions where ambipolar diffusion is very efficient at decoupling the gas from the magnetic fields. In the external parts of the disc though, the toroidal component dominates. These regions correspond roughly to an Elsasser number $\mathrm{Am}>1$, meaning that ambipolar diffusion is not the dominant process in the external parts of the disc. One can define here two disc radii: the centrifugally supported structure as defined in Sect.~\ref{disk_criteria}, and the radii at which ambipolar diffusion is not the dominant process, i.e. where the magnetic fields starts being wrapped up by rotational motions.  

\begin{figure}[t]
	\includegraphics[width=0.5\textwidth]{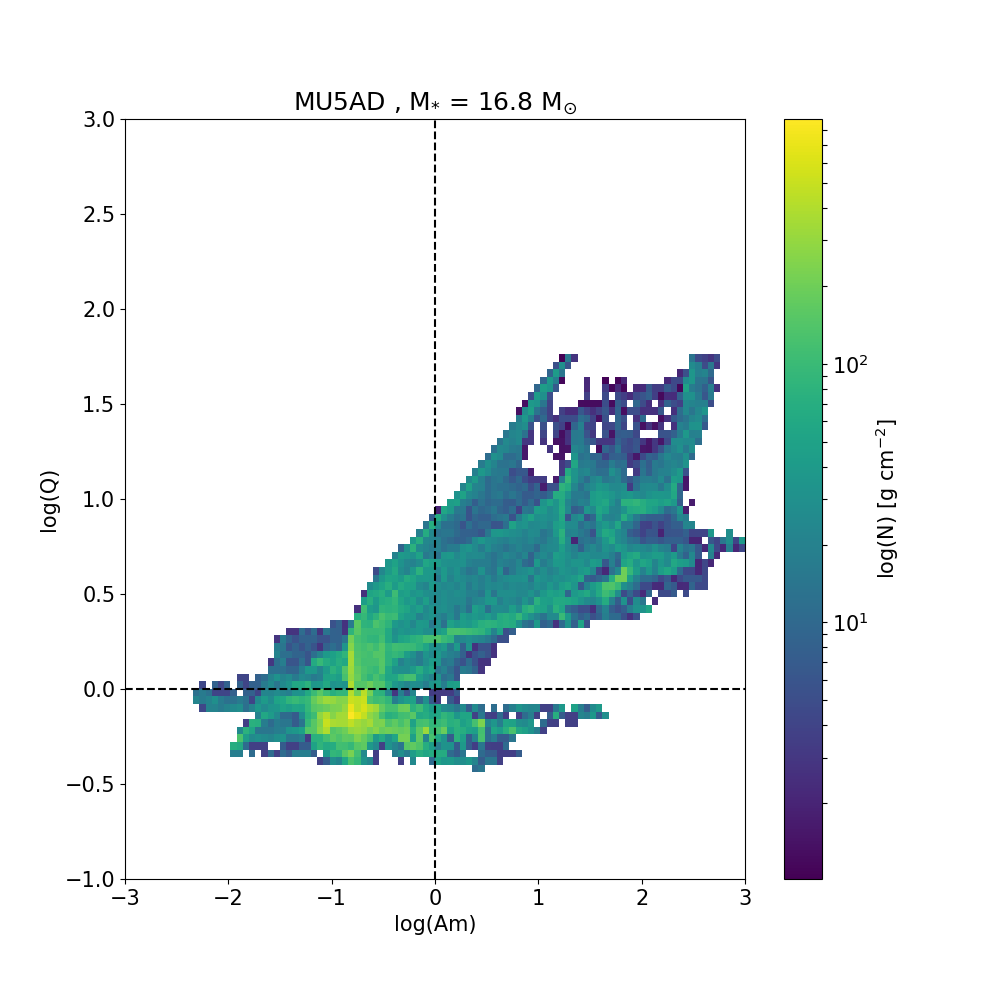}
	\caption{Histogram of the Toomre parameter $Q$  as a function of the  Elsasser number Am in the disc for the MU5AD model when the sink mass is 15~\msun. The Toomre parameter is computed as explained in the main text, as a function of the position ($r,\theta$) in the rotation plane. For each position, the Elsasser number is averaged over the disc height. The colour coding indicates the column density.  The vertical (resp. horizontal) line shows the $\rm Am=1$ (resp. $Q=1$) }
	\label{Fig:disk_AmQ}
\end{figure}

Figure \ref{Fig:disk_AmQ} shows the distribution of the Toomre parameter as a function of the Elsasser number in the disc for the MU5AD run when the sink mass is 15~\msun. We see a good correlation with increasing Toomre parameter with the Elsasser number. In particular, most of the disc material that is Toomre unstable exhibits $\mathrm{Am}<1$. On the opposite, the stable regions of the disc are mostly not dominated by ambipolar diffusion. 

To summarise, we find that the magnetic fields topology and amplitude change dramatically when resistive MHD is considered compared to the ideal MHD case. Importantly, the inner region of the discs in the resistive models are gravitationally unstable and dominated by the thermal pressure and by resistive effects, with essentially a vertical magnetic field, and an Elsasser number Am$<1$. The outer parts of the disc are gravitationally stable and exhibit a toroidal magnetic field with a large magnetic pressure dominating over the thermal one ($\beta > 1$), and Am$>1$.

\section{Discussion}\label{Sec:discussion}

In this section, we first discuss the mechanisms which regulate the disc formation and early evolution in the resistive runs. Then we compare our findings to previous work in both the low- and high-mass star formation regime, and detail the limits of our current study and numerical framework. 

\subsection{Is the disc size regulated by ambipolar diffusion?}

The disc properties we have reported so far in the previous section are dependent on the physics included as well as on the initial level of rotation. While estimating analytically the centrifugal radius in the hydrodynamic case is straightforward, it becomes more heavily physics dependent when magnetic fields are taken into account because of magnetic braking and magnetic diffusion processes. \cite{hennebelle:16} proposed a semi-analytical model to estimate the size of the disc in the case of a flow dominated by ambipolar diffusion. 

The disc radius inferred by \cite{hennebelle:16} corresponds to the radius at which the ambipolar diffusion starts to take over all other dynamical processes (induction, rotation, and freefall). It reads
\begin{equation}
R_\mathrm{AD} = 18~\mathrm{au}~\times\delta^{2/9} \left(\frac{B_\mathrm{z}}{0.1~\mathrm{G}}\right)^{-4/9}\left(\frac{\eta_\mathrm{AD}}{0.1~\mathrm{s}}\right)^{2/9} \left(\frac{M_{\star}+M_\mathrm{disc}}{0.1~\rm{M}_\odot}\right)^{1/3},
\label{Eq:disk_ana}
\end{equation}
where $B_z$ and $\eta_\mathrm{AD}$ are the vertical magnetic field amplitude and the ambipolar resistivity evaluated at the disc radius. The parameter $\delta$ corresponds to  a normalization parameter with respect to the Singular Isothermal Sphere profile \citep[SIS, ][]{shu:77}.  In Fig.~\ref{Fig:profile_rhobeta} we estimate a difference between the measured density profiles and the SIS of roughly two orders of magnitude.  Since the rotation and the magnetic pressure also participate in the establishment of the density profile  \citep[e.g.][]{hennebelle:04}, we assume that $\delta$ is on the order of a few. For the sake of simplicity, we take $\delta=1$ hereafter since the dependency of the analytic estimate scales as $\delta^{2/9}$.

\begin{figure}[t]
	\includegraphics[width=0.5\textwidth]{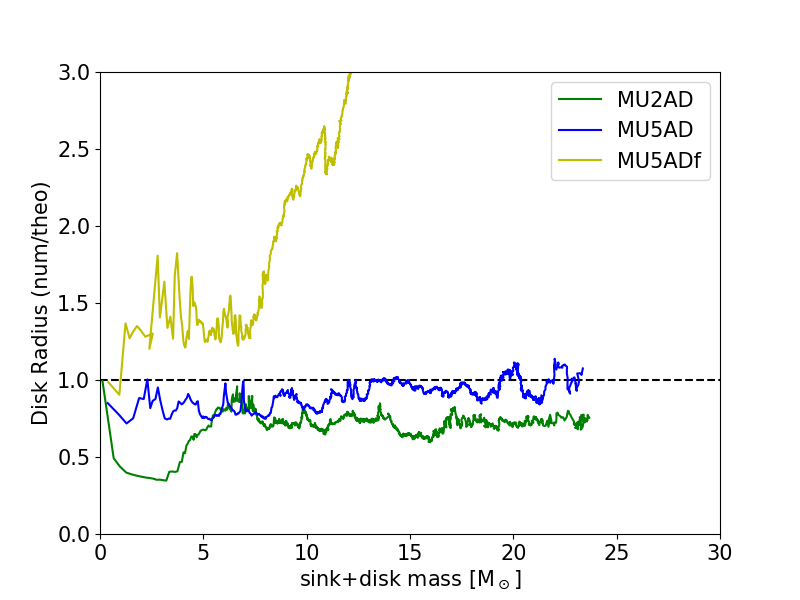}
	\caption{Ratio of the disc radius measured in the models with ambipolar diffusion (MU2AD in green, MU5AD in blue, and MU5ADf in yellow) and the analytical prediction of \cite{hennebelle:16} as a function of the total sink and disc mass. }
	\label{Fig:disk_radius_vs_analytical}
\end{figure}

Figure~\ref{Fig:disk_radius_vs_analytical} shows the evolution of the ratio of the disc radii measured in the MU2AD, MU5AD, and MU5ADf runs over the analytical prediction computed following Eq.~(\ref{Eq:disk_ana}). The agreement is within less than a factor 2 for the slow rotation models, while it increases up to a factor  larger than 3 in the fast rotating MU5ADf run at mass larger than 8~\msun. Indeed, the disc part in which the Elsasser number $\mathrm{Am}>1$ in the MU5ADf run is much larger than the one in the MU5AD one. At radii $>100-200$~au, the rotation time becomes longer than the ion-neutral collision time  ($\mathrm{Am}>1$), meaning that the magnetic fields and the neutral are well coupled (through the collisions with ions) and a toroidal component is efficiently generated by the differential rotation.  We have indeed shown in Fig.~\ref{Fig:profile_Bv} that the toroidal magnetic field component strongly dominates over a large fraction of the disc in the MU5ADf  run for sink masses $>5$~\msun. In the outer disc, the Alfv\'en velocity is also comparable to the sound speed, which indicates that the importance of the toroidal magnetic pressure support.  Equation~(\ref{Eq:disk_ana}) is valid in the region where ambipolar diffusion is efficient at preserving the generation of toroidal magnetic fields by the differential rotation. Clearly, the analytical estimate does not apply in the outer regions of the disc at late time, but rather on the inner part dominated by ambipolar diffusion.

Ambipolar diffusion thus regulates obviously the disc formation and early evolution, but then as disc grows, the ambipolar diffusion becomes negligible in the outer parts. This said, the plasma $\beta$ remains on the order of unity, meaning that the generation of toroidal field by differential rotation remains limited compared to the ideal MHD case where the toroidal fields dominate over more than one order of magnitude. 

\subsection{Characteristics of the star-disc-outflow system and observational prediction}

We summarise in this section the characteristic features we have observed in our models, depending on the physics included. We restrict our analysis to the period that we cover in this study, prior to a stellar mass of 20~\msun. 

First, in the hydrodynamical case, we have observed that the disc is the largest, the most massive, as well as the most gravitationally unstable. A very weak outflow is formed (height of a few 100s~au), in which the gas is accelerated by the radiative force.  This picture does not change qualitatively  if we account for a more realistic irradiation scheme thanks to which the radiative outflow can expand further away as a consequence of a larger radiative acceleration \citep{mignon:20}.

Second, in all our magnetised models, a system made of a star, a disc, and an outflow  is formed. For stellar mass $M_\star<20$~\msun, the outflow has a mass of 2-10~\msun~and can extend up to a few 10 000s~au. The outflow is essentially made of gas accelerated by  magnetic processes and is self-collimated by magnetic fields. The magnetic field in the outflow is thus essentially toroidal. The fact that both the ideal and resistive cases share the same properties for the early evolution of the outflow let us think that the outflow is generated in the upper layers of the disc, where ionisation is high.

For the disc, one should distinguish the ideal run from the resistive runs which exhibit opposite features. On the one hand, in the ideal case, the disc is puffy, supported by the magnetic pressure and gravitationally stable. The magnetic field is essentially toroidal throughout the disc. On the other hand, the disc in the resistive runs is thinner  and supported mostly by the thermal pressure.  One can distinguish two parts in the disc: the inner region that is gravitationally unstable where ambipolar diffusion is dominant and the magnetic field is vertical, and the outer disc  that is gravitationally stable  and where ambipolar diffusion is negligible and magnetic field is mostly toroidal.

These characteristics features can then help us to discriminate the important physical processes at play in the vicinity of massive star forming region. If a collimated outflow is observed in objects less massive than 20~\msun, then we predict that the outflow should be accelerated by magnetic processes. Models neglecting magnetic fields are unable to launch extended collimated outflows for this mass range.  Second, the size of the disc can help us to discriminate again about the importance of magnetic fields. We expect larger disc to form in the hydrodynamical case. Nevertheless, we are here discussing about factors of 2. In order to have a clearer distinction, polarised emission observations of the disc should help first to state about the presence of magnetic field, and second about the importance of resistive effect (presence of a vertical field). 

It is worth to note that the plasma beta and magnetic topology are key quantities which govern  the subsequent evolution of the disc  \citep[e.g.][]{fromang:07,flock:11,bethune:17}. Our models can help to put constraints in the initial conditions for further studies looking at the evolution of the protostellar disc as it is widely done for low-mass stars.

A step forward would be to provide synthetic observations of dust emission and polarization from our models, but this goes beyond the scope of this paper. We provide in Sect.~\ref{Sec:obs} a brief qualitative comparison between our models and recent observations.

\subsection{Comparison with previous work}
Multi-dimensional simulations of massive stars including hydrodynamics and radiative transfer have been performed since about twenty years.  \cite{yorke:02} present 2D axisymmetric calculations of the collapse of  slowly rotating massive cores of mass ranging from 30 to 120 \msun, and compare the results obtained with a grey irradiation with the ones obtained with a frequency-dependent model. They accounted for the internal luminosity of the forming protostars, as well as the accretion luminosity (total luminosity). They show that the final mass of the protostar can be increased by a factor 2 when accounting for multi-frequency radiative transfer (42.9~\msun~ against 22.9 \msun). They show that massive stars can be formed via accretion through a disc. The anisotropy of the radiative flux due to this disc is called  the flash-light effect, and is enhanced with the effects of frequency-dependent radiation transfer. \cite{kuiper:11,kuiper:12} extended this work using 3D models with a hybrid method to treat the frequency-dependent irradiation coming from the central star. They confirm the importance of the frequency-dependent irradiation as well as the flash-light effect which helps to form massive stars via disc accretion. In addition, they show that the expanding cavity is  radiation-pressure-dominated and remains stable over time, which is consistent with our results in the hydrodynamical case, although we use a grey irradiation.

\cite{krumholz:07a} performed 3D AMR RHD simulations of the collapse of 100-200 \msun~ turbulent massive cores accounting for the total luminosity of the forming protostars with a grey radiative transfer. They find that radiative feedback has a dramatic impact on the dynamics of the collapsing clouds by heating the gas and preventing it from strongly fragmenting. They report the formation of unstable discs of size $\approx 500$ au around the massive protostars which are able to channel mass inward very rapidly due to large-scale gravitational instability. Later on, \cite{krumholz:09} showed that the cavities driven by the radiative pressure may be unstable to radiative Rayleigh-Taylor   instability (RTI) which help to channel the gas onto the massive protostars through filaments that self-shield against radiation. \cite{rosen:16} extended this work using a frequency dependent irradiation scheme and confirm the development of RTI. Our results in the hydrodynamical case do not exhibit RTI development. Indeed, in a companion work, \cite{mignon:20} show that the radiative outflow cavity is  stable because the gas velocity through the interface of the cavity is large enough to prevent the development of RT instabilities if one accounts for a more accurate  irradiation scheme. We note that our results are consistent with the one of \cite{klassen:16} who present 3D AMR RHD simulations with a hybrid radiative transfer solver. They find stable bubbles expanding through radiation pressure, as well as large protostellar discs that grow rapidly and become Toomre unstable. The disc does not fragment but form spiral arms and channels material onto the star at accretion rate of a few $10^{-4}$~M$_\odot$~yr$^{-1}$ in the case of a 100~M$_\odot$ core.

Nevertheless, the picture of the outflow launching and disc formation changes dramatically when magnetic fields are taken into account. The disc is more stable and the magnetic outflow develops much earlier than the radiative one. The magnetic outflows develop quickly, i.e. before radiation dominates the acceleration, on 10000s au scale and is well collimated. The magnetic outflow thus creates a channel  in which the intense radiation of the forming protostars radiation  will efficiently escape. This result was indeed anticipated in models accounting for sub-grid protostellar outflow models on top of the protostellar irradiation, where radiation can escape easily in the outflow channel \citep[e.g.][]{krumholz:05,cunningham:11,kuiper:15}.  The theoretical framework for the development of the radiative Rayleigh-Taylor instability in presence of magnetic fields is yet missing in the literature. In addition, we should though point out that our results remain limited to the very early stages, when the central mass is $<20$~\msun. Additional work, pushing time integration toward higher masses should be carried out in order to address properly the development of the RTI.

\cite{cunningham:11} present similar AMR simulations than \cite{krumholz:07a} but including  the feedback effects protostellar outflows using a sub-grid model, on top of the protostellar radiative heating and radiation pressure exerted on the infalling gas. They show that the protostellar radiation  focused in the direction of protostellar outflow cavities is sufficient to prevent the formation of radiation pressure-supported circumstellar gas bubbles. 
\cite{kuiper:15} present 2D axisymmetric simulations including both the radiative feedback and protostellar outflow from massive stars via sub-grid models. They show that the kinematic feedback is predominant at early times, whereas the radiative acceleration becomes significant at later times. The outflows open a cavity extending to the core edge in which the radiation escapes. The outflows extend the flash-light effect from the disc scale to the core scale, and help to form more massive stars. All together, we confirm these results but the outflows we reported in this study are self-consistently launched by magneto-centrifugal processes.

To evaluate the effect of magnetic fields, \cite{seifried:11,seifried:12} present 3D MHD simulations of collapsing 100~M$_\odot$ cores in the ideal MHD approximation and neglecting the protostellar radiative feedback. They found that for weak magnetic fields ($\mu>10$) well-defined Keplerian discs with sizes of a few 100~au are formed, whereas their formation is suppressed for strong magnetic fields ($\mu<10$) due to a very efficient magnetic braking.  At first sight, this result seems contradictory with our finding in the MU5I run. We explain this by the large central mass-to-flux ratio used in our models ($\mu_\mathrm{c}=50$ for MU5 models) while they set up a non-uniform magnetic field with a larger amplitude in the centre than ours ($>500$~G versus $\simeq 70$~$\mu$G). The accretion rates they observed are of the order of a few $10^{-4}$~M$_\odot$~yr$^{-1}$. The outflows observed in their simulations are launched by magneto-centrifugal acceleration, which are initially poorly collimated, and then get better collimated over time due to the development of fast jets.
On the disc scales, \cite{myers:13} perform 3D AMR ideal MHD simulations of collapsing turbulent and magnetised 300~\msun~ massive cores, with a maximum resolution of 10 au. They include radiative transfer with a grey FLD approximation, as well as radiative protostellar feedback. They report the formation of Keplerian discs, with a radius of 40 au when the sink mass is 3.5~\msun. In the MU5I model, we measured a disc size of $\simeq75$~au at the same sink mass, which is fairly similar given all the differences between \cite{myers:13} numerical setup and ours. Interestingly, they report the presence of episodic outflows of velocities $\simeq10$~km~s$^{-1}$, which get stronger once the sink mass gets larger than 20~\msun. The outflow velocity, as well as the episodic feature, are very consistent with our findings in the MU5I model. Altogether, our results using the ideal MHD approximation compare well with the literature. This strengthens the importance of resistive effects on the early evolution of discs and outflows in young massive protostars. 

\cite{matsushita:17} study 3D collapse using resistive MHD  (Ohmic resistivity) and a barotropic equation of state to  mimic the thermal behaviour of the collapsing gas. For strong magnetic fields, they find that magnetically driven massive outflows are launched, whereas they are subdued or absent for weak magnetic fields. In  addition, in the weakly magnetised case, they show that  fragmentation occurs which prevents the formation of massive star. The outflows they report have a wide opening angle at the disc scale and a collimated structure at large scales, similar to what we have presented in this study.
\cite{kolligan:18} present 2D axisymmetric simulations of the collapse of 100~\msun~ dense cores  using Ohmic diffusion and an isothermal equation of state. They have a higher resolution than ours ($\Delta x=0.09$~au, sink cell size of  1 au) and integrate up to a central sink mass $>50$~\msun. They report disc and outflow formation, in a similar qualitative pictures than the one presented in the resistive runs of this study. The disc size they report is larger than ours ($>1000$ au), but given the differences in numerical methods (dimensionality, grid) and physics included (Ohmic versus ambipolar diffusion) it is not clear what is the main reason of this difference. Further comparison work is required. For the outflow, \cite{kolligan:18} report the formation of a  magneto-centrifugally launched, highly collimated central jet and  a slow wide-angle magnetic-pressure-driven tower flow. The latter component is launched in the outer disc and dominates the angular momentum transport, similarly to what we find in Fig. \ref{Fig:angmom}. We note that we do not retrieve the high-velocity jet component reported by \cite{kolligan:18} in our results, which is certainly due to our limited numerical resolution that does not allow to reach large rotation velocity. Indeed, the ejection velocity is linked to the (Keplerian) rotation velocity in the standard theory of outflow launching \citep[e.g.][]{spruit:96}.
While we did not investigate the exact origin of the magnetic outflow (magnetic tower or magneto-centrifugal acceleration), we confirm that even in the presence of radiative feedback, the development of magnetised outflows is ubiquitous as first demonstrated by \cite{banerjee:06}. We refer readers to \cite{mignon:21b} for a comprehensive study of the outflows in similar models. For the disc, our results confirm that disc can form, even in the ideal MHD regime. We are the first to study in detail its structure depending on the physics taken into account (magnetic fields topology, plasma beta). In particular, we demonstrate that the results of low-mass models can be extended to higher mass and that resistive effect change completely the properties of the disc (aspect ratio, stability). Further work is needed, as for instance the inclusion of the two other resistive effects, the Ohmic diffusion and the Hall effect, which are found to influence disc formation in the low-mass regime \cite[e.g.][]{tsukamoto:15,vaytet:18,marchand:19,zhao:20}. 

Our centrally condensed initial conditions, which are similar to the choice made in other work \citep{krumholz:09, seifried:11,mignon:21a},  do not favour fragmentation. Our results regarding fragmentation are thus strongly biased by this choice and only disc fragmentation is expected to occur. The fragmentation we report in this study is consistent with previous studies. In the HYDRO run, a binary system is formed from disc fragmentation with most of the mass going in the primary fragment, similarly to what has been reported in \cite{krumholz:07a}. In the ideal MHD case, we did not report fragmentation  because of the extra support provided by magnetic fields. This results is consistent with the one of  \cite{seifried:11} for strong field case.  For the resistive runs, the discs are gravitationally unstable in the inner regions but do not fragment. Similar results are reported in \cite{matsushita:17} and \cite{mignon:21a} in the aligned case. We note that \cite{mignon:21a} report disc fragmentation in the case where initial turbulence is super-Alfvénic. Regarding fragmentation, further time integration is also required in order to investigate the disc stability evolution as the central star gets more massive.  

 Last, we did not take into account turbulence nor misalignment in our initial setup whereas  numerical experiments in the literature have proven that they affect the magnetic braking efficiency and enables the formation of Keplerian discs \citep[e.g.][]{hennebelle:09,santos:12,joos:12,joos:13,seifried:13}. We present in \cite{mignon:21a,mignon:21b} an extension of the present models with ambipolar diffusion where we include turbulence in the initial setup, as well as a more accurate scheme for stellar irradiation based on \cite{mignon:20}. We show that the disc properties remain unchanged because ambipolar diffusion dominates in the inner disc. Regarding fragmentation, sub-Alfvénic models of \cite{mignon:21a} do not fragment as the ones in the present study, but  super-Alfvénic models lead to disc fragmentation and binary formation. Interestingly, the properties of the discs surrounding each star in the binary system remain very similar to what we report here. Regarding outflows, \cite{mignon:21b} show that the  outflow remains magnetically driven at early stages in sub-sonic models, even if  the contribution of the radiative acceleration gets larger thanks to the hybrid irradiation scheme which better treats stellar photons. We note however that the outflow launching is strongly perturbed in the case of initial supersonic turbulence.

\subsection{Comparison with observations \label{Sec:obs}}

Massive star formation regions are usually observed at distances larger than a kpc, which limits the resolution of such observations. Very few disc candidates in young massive star forming regions have thus been reported in the literature \citep[see][for a review]{beltran:16,rosen:20}. We review a few of them in the following section and compare with our results. 

\cite{ahmadi:18}  study the fragmentation and kinematics of the high-mass star-forming region W3(H$_2$O) and found indications for possible disc fragmentation on 1000 AU scales. In all our model, we do not report such wide fragmenting disc. The most favourable case to compare with this work would be the HYDRO case because of the large disc radius, but it does not reproduce the observed highly collimated and massive (10~\msun) outflows reported by \cite{zapata:11} in W3(H$_2$O). We must stress though that our initial conditions are not chosen to be favourable to disc fragmentation because of the low initial rotation level. The influence of initial turbulence should also be investigated in that purpose. 

More recently, \cite{motogi:19} reported ALMA observation of a young high-mass protostellar object (10~\msun, no ultra-compact HII region), accreting at about $10^{-3}$~\msun~yr$^{-1}$, which characteristics are very consistent with the early evolutionary scenario of a low-mass protostar. From dust continuum emission, they report a disc mass of $2-7$~\msun~and a disc size of $250$ au, associated with a lower limit of 0.4 for the Toomre parameter. Altogether, this observation is very consistent with our models including ambipolar diffusion though their estimated protostellar age ($10^4$~yr) is relatively shorter. 

\cite{patel:05} found in Cepheus A HW2 a rotating disc-like structure of mass 1-8~\msun~and size 330 au centred on a massive 15~\msun~protostar . \cite{fernandez:16} reported polarised emission observations from this disc-like structure and found indication of either a uniform magnetic field threading the disc (polarised emission) or of grain growth up to a few 10s $\mu$m sizes within a few $10^4$~yr (scattering). They exclude the possibility of polarised emission coming from a toroidal field. On top of this, \cite{vlemmings:06} measured magnetic field strengths in the HW2 disc area ranging from 100 to 600~mG. All together, this object is again very consistent with the models integrating ambipolar diffusion we present here:   a moderate magnetisation at the disc border (10-100~mG) and a rather uniform vertical field in the inner disc, contrasting with the stronger magnetisation and the toroidal field found in the ideal MHD case. 

 Last, a couple of recent observations report Keplerian motions associated with compact sources (discs?) in young massive protostars thanks to the ALMA interferometer. \cite{maud:19} present  observations of the G17.64+0.16 young massive protostar ($\simeq 45~$\msun) which revealed a disc of size $\simeq 120$ au associated with Keplerian rotation. Similarly, \cite{ginsburg:18} report Keplerian rotation and a disc size of 50 au around the Orion SrCI source ($\simeq 15$~\msun). These observations are consistent with the disc size we report in the MU2AD and MU5AD runs.  \cite{johnston:15,johnston:20} report near-Keplerian rotation associated with a disc  size of  1000 au in the massive protostar AFGL4176mm1. The disk exhibits substructures (possibly spiral arm)  and its stability analysis shows that it is  gravitationally unstable. Only the HYDRO run in our models can lead to such large disc radius. However, \cite{mignon:21a}  show that if initial supersonic and super-Alfv\'enic turbulence are considered, such large rotating and unstable structures can form in magnetised models with ambipolar diffusion (see their figure 8). 

This qualitative comparison remains highly biased toward high resolution observations of massive star disc candidates which allow to directly probe and resolve the disc scales in order to compare with our models. A more careful and quantitative analysis, including synthetic observations for a side by side comparison, is clearly required but this goes beyond the scope of the current paper.

\subsection{Limits of the model and future work\label{Sec:discussion_limits}}

We have identified two types of limits which may affect the generalization of our conclusions. The first comes from the numerical methods we used, and the second is about the initial conditions and parameter space exploration. 

First we use an aggressive merging scheme for the sink particles, meaning that every sink particle that enters the accretion radius of another more massive one gets immediately merged. We choose on purpose to merge all the overlapping sink in order to favour mass accretion and the formation of massive stellar sources. We have reported the formation of secondary sink particles in the HYDRO run, but most secondary are merged with the central one  within a few hundred years and the mass of the accreted secondary sink particles remains small compared to the central one (more than a factor 10 in mass). The HYDRO run ends in a binary system and a secondary sink of mass $\simeq 0.01$~\msun~ gets ejected from the disk. In all magnetised models, the system ends in a single star without secondary sink formation. 
The second limitation comes from the grey FLD irradiation scheme we use. It has been shown in the literature that averaging over frequencies and using only the zeroth-moment of the radiative transfer equation can underestimate by two orders of magnitude the radiative force \citep[e.g.][]{kuiper:12}.  Besides, the isotropic FLD irradiation scheme is also expected to have an effect on the disc radial temperature profile, which might then affect the disc stability. 
\cite{mignon:20} propose in a recent methodological paper an improvement of the irradiation scheme using the M1 moment models to handle stellar irradiation \citep{rosdahl:13,rosdahl:15}. We note that the recent work of \cite{mignon:21a,mignon:21b} investigate the effect of a better irradiation scheme combined with ambipolar diffusion and initial turbulence. They confirm qualitatively our results,  in particular on the relative importance of the magnetic versus radiative accelerations up to $M_\star\simeq 20$~\msun, and  on the magnetic properties of the discs.

	Third, we do not explore the influence of the fraction of the incident kinetic energy radiated away, i.e., the accretion luminosity. This choice is motivated b y the large uncertainties that remain in order to propose a coherent model to set the accretion luminosity.  Recent radiation-(magneto)hydrodynamics  simulations of protostellar core formation indicate that the accretion shock onto the protostar at the early stages is a subcritical radiative shock, meaning that all the incident kinetic energy  is transferred to the protostar and not radiated away \cite{vaytet:13,vaytet:18,bhandare:20}. This result applies for the very early stages of protostellar core formation, and the extrapolation of this properties to PMS evolution is very uncertain. Indeed, \cite{baraffe:09} computed PMS evolution of young low-mass protostars (up to 1~\msun) and showed that an evolution including the effects of episodic accretion at the stellar scale can explain the observed luminosity spread in H–R diagrams of star-forming regions, provided that the accretion is cold, meaning that a significant fraction of the accretion energy is radiated away. This results need to be extended to massive young protostars though. Besides \cite{hennebelle:20b} show that including the instantaneous full accretion luminosity provides strong heating so that the measured sink mass function in numerical experiment becomes top-heavy and is not agreement with the observed stellar initial mass function.  Last, the internal luminosity from the massive young protostar is expected to exceed the accretion luminosity for $M_\star\gnsim 8$~\msun ~\citep[e.g.][]{hosokawa:09}, which is the regime we target in this study. Given the large uncertainties, we do not consider any accretion luminosity here and postpone the exploration of its influence in future studies. In addition,  further work should deserve a detailed study on the way to properly account for the energy radiated away down to protostellar scales, but this goes far beyond the scope of the present study.

Then, we only explore a narrow range of initial conditions: aligned rotator (magnetic fields and rotation axis are parallel) and only two magnetization and rotation levels. It has been shown in the context of low-mass star formation that misalignment and turbulence greatly affects the formation of protostellar disc \citep{hennebelle:09,joos:12,santos:12,joos:13,masson:16}. Besides, the initial density profile may also drastically affect the fragmentation of the collapsing massive cores \citep{girichidis:11,lee:18}. The parameter space exploration will be performed in future works in order to test the resilience of our findings on the mechanisms governing the accretion and ejection processes in young massive stellar objects.

Last, we assume that dust and gas are both dynamically and thermally perfectly coupled. It has been recently shown in numerical simulation work that dust and gas may decouple dynamically in collapsing dense cores with efficient dust enrichment for grains larger than 10~$\mu$m \citep{bate:17,lebreuilly:19}. There are also growing observational evidences of large grains in the vicinity of collapsing cores, both in the low- and high-mass regimes \citep{fernandez:16,sadavoy:18,galametz:19,valdivia:19}. In addition, dust is the main opacity source so that protostellar irradiation primary couples with the dust and then to the gas via the drag. Besides, \cite{thiem:21} recently suggested that  the effects of grain rotational disruption by radiative torques in young massive protostars can lead to the destruction of  micron dust grains which results in a reduction of radiation pressure opacity. Last but not least, dust growth is expected to change quantitatively the non-ideal MHD resistivities, which may also amplify the effect of ambipolar diffusion for instance  \citep{zhao:16, guillet:20}. 
Future studies following the differential dynamics of the dust and gas mixture, coupled with magnetic fields evolution and radiation transport,  is naturally the next step forward.

\section{Conclusion\label{Sec:conclusion}}

We have presented a first suite of 3D numerical simulations integrating the combined effect of protostellar evolution and ambipolar diffusion in the context of the early evolution of  massive protostars (up $M_\star\simeq 20$~\msun). We have compared the effect of magnetic fields with respect to a pure hydrodynamical case. Then, we explore the impact of the physics (ideal versus non-ideal MHD) and of the initial conditions (magnetization and rotation level) on the formation of the star-disc-outflow system. 

First, we find that the magnetised models differ dramatically from the hydrodynamical one, with different mass accretion and particularly mass ejection rates. A disc is formed in all our models, but again with quantitative differences on the mass and radius between hydrodynamical and MHD models. More importantly, the magnetic properties of the disc formed in the non-ideal MHD framework are opposite compared to ideal MHD. While the latter case exhibits strong toroidal fields throughout the disc, with plasma $\beta<1$, the non-ideal MHD inner discs are dominated by the thermal pressure ($\beta>1$). In addition, these discs are threaded by a vertical field in their inner parts ($\leq 100-200$~au) and generate a toroidal field in the outer parts via differential rotation. Correspondingly, the inner regions of the disc in the resistive runs  are gravitationally unstable and dominated by resistive effects, while the outer parts are gravitationally stable and more coupled to the magnetic fields evolution. This result, as well as the disc radius, can be tested in future high angular resolution observations. It also puts well-defined constraints on the initial conditions for the subsequent evolution of (protoplanetary?) discs around massive stars. 

Second, we find that the accretion-ejection is regulated by the formation of discs and magnetised outflows which dominates mass ejection at the early stages.  Our results confirm that when outflows are self-consistently launched, the well-studied  flash-light effect (outflows create a channel for the radiation to escape at large distances) still holds. 

Magnetic fields and ambipolar diffusion  affect dramatically  mass accretion and ejection and disk formation in the early stages of massive star formation. Our analysis of the accretion/ejection mechanisms at play in young massive protostar are a scale-up version of the one occurring within low-mass protostars. The exploration of the initial parameter space yet deserves future works in order to test the robustness of the accretion/ejection magnetically regulated scenario, in particular in the presence of initial turbulence. The effect of a more accurate irradiation scheme as well as the differential dynamics of dust and gas has to be investigated in the near future.   Besides, our results remain limited to the early stages of  (very) massive star formation, and further works need to be carried out for stellar mass $>20$~\msun.

\begin{acknowledgements}
We thank Ugo Lebreuilly and Rolf Kuiper for useful discussion. We thank Rolf Kuiper for kindly sharing the PMS evolution tracks. We acknowledge financial support from "Programme National de Physique Stellaire" (PNPS) of CNRS/INSU, CEA and CNES, France. NV gratefully acknowledges support from the European Commission through the Horizon 2020 Marie Sk{\l}odowska-Curie Actions Individual Fellowship 2014 programme (Grant Agreement no. 659706).
This work was granted access to the HPC resources of CINES (Occigen) under the allocation 2018-047247 made by GENCI.  Part of the simulations were performed at the PSMN (P\^ole Scientifique de Mod\'elisation Num\'erique) of the ENS de Lyon. We additionally acknowledge support and computational resources from the Common Computing
Facility (CCF) of the LABEX Lyon Institute of Origins (ANR-10-
LABX-66). In addition, we thank the Département d'Astrophysique, IRFU, CEA Saclay, and the Laboratoire Astrophysique Instrumentation Mod\'{e}lisation, France, for granting us access to the supercomputer \texttt{IRFUCOAST-ALFVEN} where the groundwork with many test calculations were performed.
Figures 1, 2, 7, 8, 9, 12, and B.2  were created using the \href{https://github.com/nvaytet/osyris}{\texttt{OSYRIS}}\footnote{\url{https://github.com/nvaytet/osyris}} visualization package for \texttt{RAMSES}. Figures \ref{Fig:morpho1} was created using  the VisIt\footnote{\url{https://wci.llnl.gov/simulation/computer-codes/visit}} software.
\end{acknowledgements}

\bibliographystyle{aa}
\bibliography{biblio}

\begin{appendix}

\section{Resolution convergence}
\label{Sec:appendix_resolution}

In this section, we discuss the influence of the spatial resolution on our results. We run three new models similar to MU5AD but changing the maximum level of refinement $\ell_\mathrm{max}$ of the AMR grid. MU5AD has a $\ell_\mathrm{max}=15$ corresponding to a finest resolution of 5~au. We run simulations with $\ell_\mathrm{max}=13$, $\ell_\mathrm{max}=14$, and $\ell_\mathrm{max}=16$, corresponding to finest resolution of 20~au, 10~au, and 2.5~au respectively.   The sink accretion radius varies with resolution and equals four times the finest resolution. Due to the computational cost, the simulation with $\ell_\mathrm{max}=16$ cannot be run until long times and has only reached a sink mass of 7~\msun.

\begin{figure}[t]
	\includegraphics[width=0.5\textwidth]{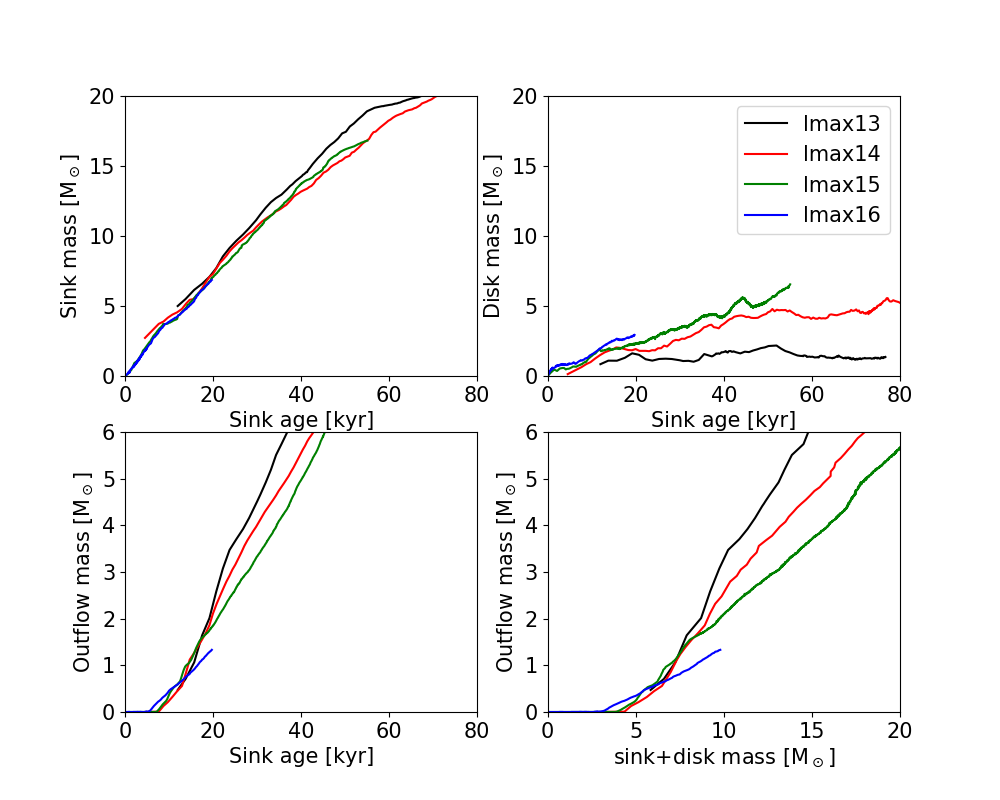}
	\caption{Mass evolution of the sink (top-left), disc (top-right), outflow (bottom-left) as a function of time after sink creation, and of the outflow as a function of the total disc+sink mass (bottom-right). The different colours scales for different maximum level of refinement $\ell_\mathrm{max}$ and resolution: 20~au (black), 10~au (red), 5~au (green), and 2.5~au (blue).}
	\label{Fig:mass_resolution}
\end{figure}

Figure~\ref{Fig:mass_resolution} shows the evolution of the sink, disc, and outflow for these four runs. The sink mass as a function of the sink age (upper left panel) appears to be independent of the resolution. Concerning the disc, its mass increases with time in all models but the lowest resolution one  $\ell_\mathrm{max}=13$. The global trend is a increasing disc mass with increasing the resolution.  As for the outflow, its mass increases in time for each simulation, but its growth rate is larger at coarser resolution (opposite behaviour than for the disc mass). Interestingly, we measure outflow velocities which increase with the resolution. This further confirms the magnetic origin of the outflow, which velocity scales with the rotation velocity at its base. 
The fiducial value  used in this paper ($\ell_\mathrm{max}=15$, 5~au) seems not be converged quantitatively for the disc and outflow mass. But finest resolution for all the runs are computationally out of reach and the qualitative results are the similar whatever the resolution. We therefore think that our qualitative conclusions stand valid.

\section{Dependence on the angular momentum sub-grid model}
\label{Sec:appendix_faccmom}
In this section, we discuss the influence  on our results of the angular momentum accretion in the sink algorithm. In our fiducial model MU5AD, all the angular momentum of the gas is accreted onto the sink. We run a new model where the angular momentum is not accreted.

\begin{figure}[t]
	\includegraphics[width=0.5\textwidth]{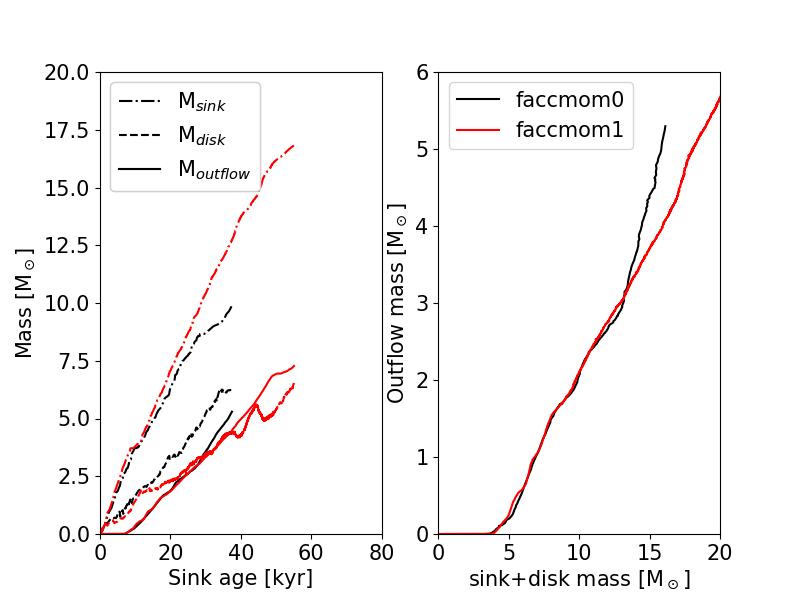}
	\caption{Mass evolution as a function of the sink age (left panel) or the total mass (right panel) for the models with (red) or without (black) the angular momentum accreted}
	\label{Fig:mass_faccmom}
\end{figure}

\begin{figure}[t]
	\includegraphics[width=0.5\textwidth]{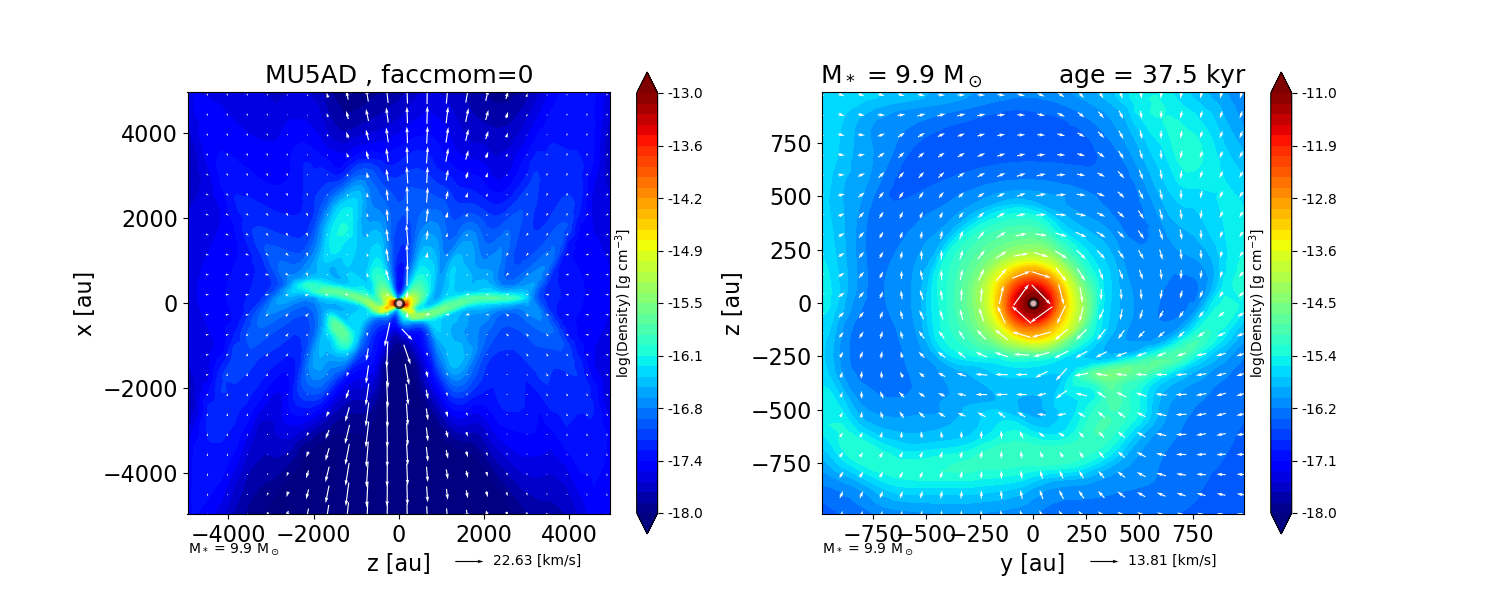}
	\includegraphics[width=0.5\textwidth]{density_zoom2_00498}
	\caption{Density maps in the plane of the disc and perpendicular to it for the model with no angular momentum accreted (top), and with all the angular momentum accreted (fiducial model MU5AD, bottom)}
	\label{Fig:densitymaps_faccmom}
\end{figure}

Figure ~\ref{Fig:mass_faccmom} shows the mass evolution of the different components (sink, outflow, and disc) in the two simulations. The sink mass grows  faster when the angular momentum is accreted. Indeed, in that case, the gas around the sink particle lowers its angular momentum and then falls more easily on the sink. On the contrary, the rotationally supported disc is favoured in the case where the angular momentum is not removed. For the same reason, the outflow, as it is generated by magnetic processes, is also more massive and stronger with no angular momentum accreted, as it can be seen also in Fig.~\ref{Fig:densitymaps_faccmom} which shows the density maps in the disc plane and perpendicular to it when the sink has a mass of 10~\msun. The morphologies are very similar, even if the disc is larger when angular momentum is not accreted.

\end{appendix}

\end{document}